\documentclass[acmsmall, screen]{acmart}

\AtBeginDocument{%
  \providecommand\BibTeX{{%
    \normalfont B\kern-0.5em{\scshape i\kern-0.25em b}\kern-0.8em\TeX}}}

\setcopyright{none}

\acmConference[CSCW'25]{2025 Conference on Computer Supported Cooperative Work and Social Computing}{October 18--22, 2025}{Bergen, Norway}
\usepackage{tabularx} % for tabularx environment
\usepackage{booktabs}
\usepackage{longtable}
\usepackage{xcolor}
\PassOptionsToPackage{table}{xcolor}
\usepackage{graphicx}
\usepackage{subcaption}
\usepackage{wrapfig}
\usepackage{adjustbox}
\usepackage{float}
\usepackage{pdfpages}

\definecolor{burntorange}{RGB}{204, 85, 0} % Define burnt orange color

\newif{\ifhidecomments}
\hidecommentstrue
\ifhidecomments
    \newcommand{\sachinedit}[1]{#1}
\else 
    \newcommand{\sachinedit}[1]{\textbf{\textcolor{burntorange}{#1}}}
\fi

\begin{document}

\title{The Role of Partisan Culture in Mental Health Language Online}

\author{Sachin R. Pendse}
\affiliation{%
  \institution{Northwestern University}
  \city{Chicago}
  \state{IL}
  \country{USA}
}
\email{sachin.r.pendse@northwestern.edu}

\author{Ben Rochford}
\affiliation{%
  \institution{Duke University}
  \city{Durham}
  \state{NC}
  \country{USA}
}
\email{ben.rochford@duke.edu}

\author{Neha Kumar}
\affiliation{%
  \institution{Georgia Institute of Technology}
  \city{Atlanta}
  \state{GA}
  \country{USA}
}
\email{neha.kumar@gatech.edu}

\author{Munmun De Choudhury}
\affiliation{%
  \institution{Georgia Institute of Technology}
  \city{Atlanta}
  \state{GA}
  \country{USA}
}
\email{munmund@gatech.edu}

\renewcommand{\shortauthors}{Pendse et al.}

\begin{abstract}
The impact of culture on how people express distress in online support communities is increasingly a topic of interest within Computer Supported Cooperative Work (CSCW) and Human-Computer Interaction (HCI). In the United States, distinct cultures have emerged from each of the two dominant political parties, forming a primary lens by which people navigate online and offline worlds. We examine whether partisan culture may play a role in how U.S. Republican and Democrat users of online mental health support communities express distress. We present a large-scale observational study of 2,184,356 posts from 8,916 statistically matched Republican, Democrat, and unaffiliated online support community members. We utilize methods from causal inference to statistically match partisan users along covariates that correspond with demographic attributes and platform use, in order to create comparable cohorts for analysis. We then leverage methods from natural language processing to understand how partisan expressions of distress compare between these sets of closely matched opposing partisans, and between closely matched partisans and typical support community members. Our data spans January 2013 to December 2022, a period of both rising political polarization and mental health concerns. We find that partisan culture does play into expressions of distress, underscoring the importance of considering partisan cultural differences in the design of online support community platforms.
\end{abstract}


\settopmatter{printacmref=false} % hides the ACM Reference Format block
\setcopyright{none}              % disables copyright and licensing footer
\acmConference[CSCW’25]{CSCW’25}{October 18--22, 2025}{Bergen, Norway}
\acmYear{}
\acmISBN{}
\acmDOI{}
\acmPrice{}
\maketitle

\definecolor{DemocratBlue}{RGB}{21, 96, 130}  % Example RGB for HTML color 156082
\definecolor{RepublicanRed}{RGB}{192, 0, 0}   % Example RGB for HTML color C00000
\definecolor{UnaffiliatedBrown}{RGB}{112,56,24}
\definecolor{myred}{HTML}{FF0000}
\definecolor{myblue}{HTML}{156082}
\definecolor{mypurple}{HTML}{7030A0}
\definecolor{myorange}{HTML}{E97132}
\newcommand{\republicanword}[1]{\textcolor{myred}{\textit{#1}}}
\newcommand{\republicanwordnoitalics}[1]{\textcolor{myred}{#1}}
\newcommand{\democratword}[1]{\textcolor{myblue}{\textit{#1}}}
\newcommand{\democratwordnoitalics}[1]{\textcolor{myblue}{#1}}
\newcommand{\unaffiliatedword}[1]{{\color{UnaffiliatedBrown} \textit{#1}}}
\newcommand{\unaffiliatedwordnoitalics}[1]{{\color{UnaffiliatedBrown} #1}}
\newcommand{\similarword}[1]{\textcolor{mypurple}{\textit{#1}}}
\newcommand{\similarwordnoitalics}[1]{\textcolor{mypurple}{#1}}
\newcommand{\dissimilarword}[1]{\textcolor{myorange}{\textit{#1}}}

\newcommand{\matchingfigure}{
    \begin{figure}
    \includegraphics[width=.6\textwidth, height=0.39\textheight, keepaspectratio]{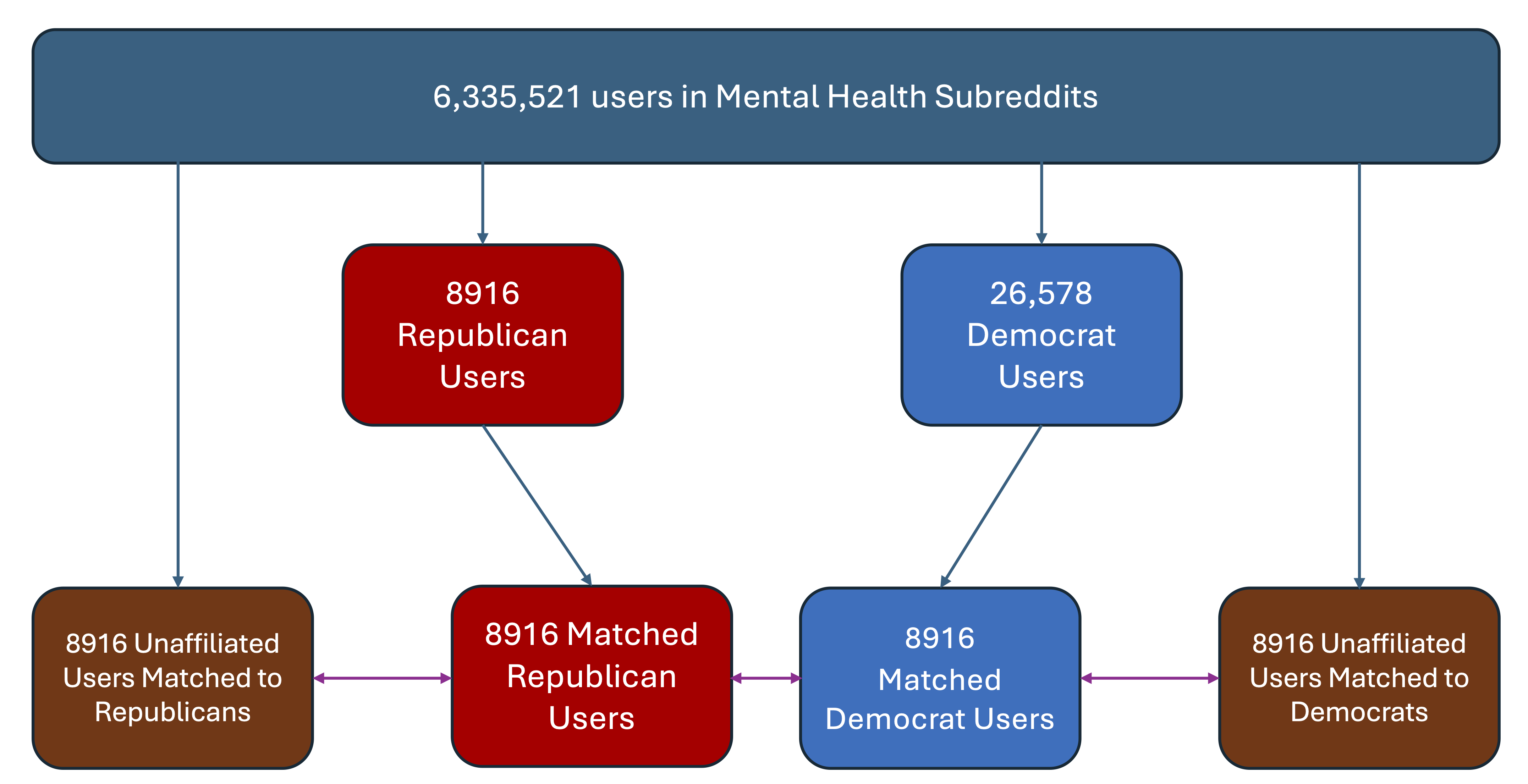}
    \caption{Between January 2013 to December 2022, 6,335,521 users posted in mental health subreddits. Of those users, utilizing Rajadesingan et al.~\cite{rajadesingan2021political} method of identifying partisans, we find 8,916 Republican Reddit users and 26,578 Democrat users. We then find the most closely matched 8,916 {\color{DemocratBlue}Democrat} and {\color{RepublicanRed}Republican} users, and match those users to separate sets of 8,916 Unaffiliated users, drawn from two 1\% random samples that exclude partisan users.}
    \label{fig:matching_figure}
\end{figure}
}

\newcommand{\analysisfigure}{
    \begin{figure}
    \includegraphics[width=.75\textwidth, height=0.39\textheight, keepaspectratio]{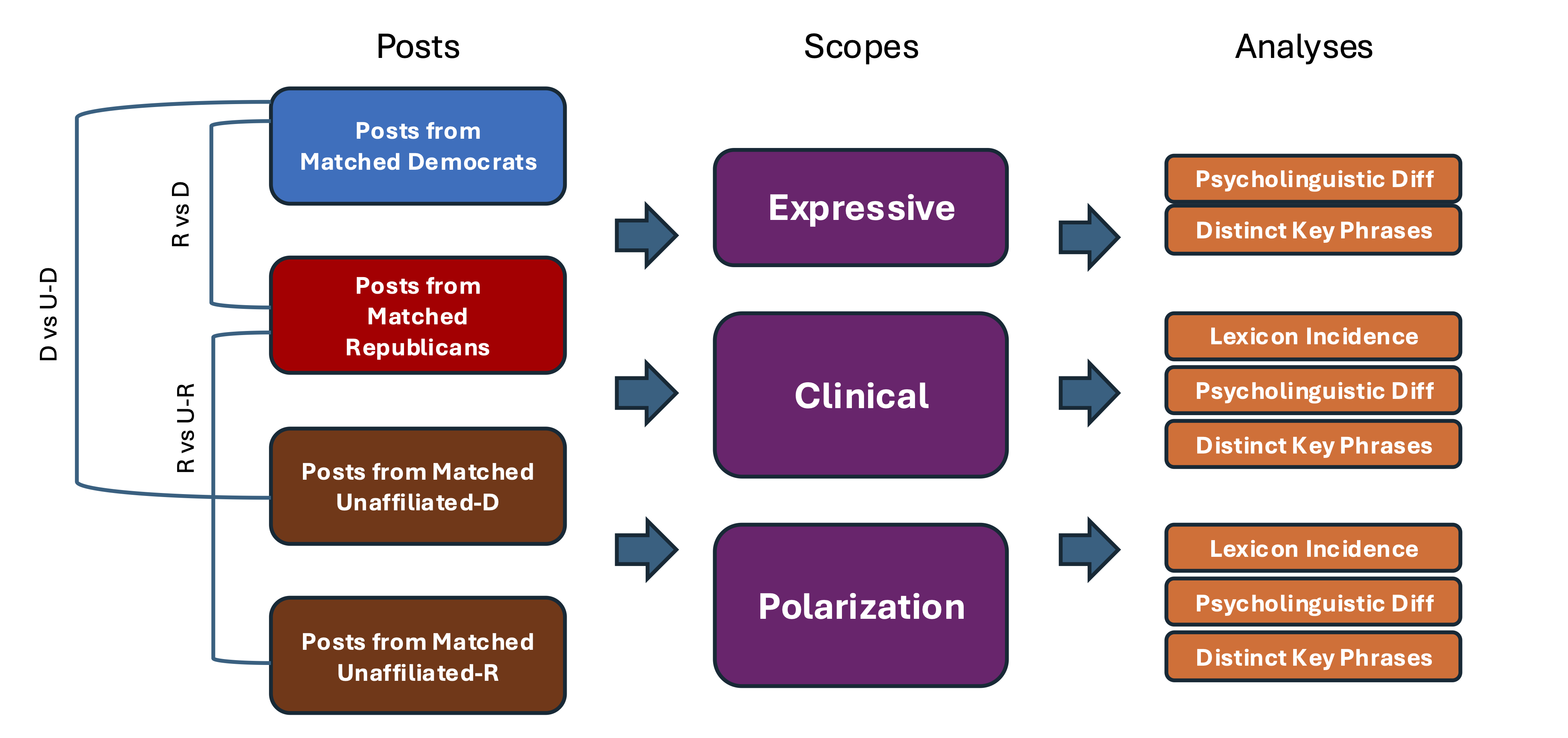}
    \caption{For our analysis, we conduct comparative analyses between matched {\color{RepublicanRed}Republican} and {\color{DemocratBlue}Democrat} users, {\color{RepublicanRed}Republican} and matched {\color{UnaffiliatedBrown}Unaffiliated} users, and {\color{DemocratBlue}Democrat} and matched {\color{UnaffiliatedBrown}Unaffiliated} users. We then evaluate differences between each group's posts, between each group's posts that use clinical language (among matched users who used clinical language), between each group's posts that use polarization-related language. Using the LIWC text analysis tool, we measure psycholinguistic differences, and using SAGE, we identify key phrases that are distinct to each population. For our clinical and polarization scopes of analysis, we also evaluate whether there are significant differences in each group's use of that specific type of language (via clinical and polarization language lexicons).}
    \label{fig:analysis_figure}
    \vspace{-5pt}
    \end{figure}
}

\newcommand{\matchingtable}{
\begin{table}[htbp]
\centering
\tiny % or \tiny
\renewcommand{\arraystretch}{0.8} % Reduces vertical padding
\begin{tabularx}{\textwidth}{lXXXXXXXr} % Added two more 'X' for the new columns
\toprule
 Analysis Type & SMD $\pm$ Std. Dev.  & AVR & Spearman $\rho$ & Kendall's $\tau$ & \# of Group 1 Posts & \# of Group 2 Posts & Matched Pairs & Total Posts \\
\midrule
\multicolumn{9}{l}{{\color{DemocratBlue}Democrat} and {\color{RepublicanRed}Republican} users:} \\ % Updated the number to 9
\midrule
 Expressive Differences & $.02 \pm .025$ & $1.24$ & $.96$ & $.84$ & 758,358 & 719,007 & 8,916 & 1,477,365 \\
 Clinical Differences & $.03 \pm .029$ & $1.26$ & $.96$ & $.84$ & 119,910 & 123,687 & 6,028 & 243,597 \\
 Polarization Differences & $.02 \pm .026$ & $1.26$ & $.96$ & $.84$ & 268,727 & 265,422 & 7,840 & 534,149 \\
\midrule
\multicolumn{9}{l}{{\color{RepublicanRed}Republican} and \unaffiliatedwordnoitalics{Unaffiliated Users}} \\ % Updated the number to 9
\midrule
 Expressive Differences & $.05 \pm .02$ & 1.17 & .95 & .83 & 758,358 & 373,261 & 8,916 & 1,131,619 \\
 Clinical Differences & $.06 \pm .02$ & $1.17$ & $.96$ & $.84$ & 117,218 & 73,807 & 5,727 & 191,025 \\
 Polarization Differences & $.05 \pm .02$ & $1.17$ & $.95$ & $.82$ & 266,729 & 131,399 & 7,619 & 398,128 \\
\midrule
\multicolumn{9}{l}{{\color{DemocratBlue}Democrat} and \unaffiliatedwordnoitalics{Unaffiliated Users}} \\ % Updated the number to 9
\midrule
 Expressive Differences & $.04 \pm .02$ & 1.17 & $.97$ & $.86$ & 719,007 & 334,183 & 8,916 & 1,053,190 \\
 Clinical Differences & $.06 \pm .02$ & $1.19$ & $.97$ & $.86$ & 121,618 & 66,639 & 5,730 & 188,257 \\
 Polarization Differences & $.05 \pm .02$ & 1.19 & $.97$ & $.86$ & 262,636 & 118,505 & 7,529 & 381,141 \\
\bottomrule
\end{tabularx}
\vspace{10pt}
\caption{Matching and dataset statistics for each user group comparison. We find close matches for all of our comparisons, using the Hungarian method to find the most optimally matched pairs via Mahalanobis distance. We also use Spearman's $\rho$ and Kendall's $\tau$ rank correlation comparisons to demonstrate that the subreddits posted in by these groups of users are well-matched as well. All Spearman's $\rho$ and Kendall's $\tau$ correlation comparisons have a significance of $p < 10^{-22}$ or lower.}
\label{tab:matching_statistics}
\vspace{-15pt}
\end{table}
}

\newcommand{\expressivedifferencesliwc}{
\begin{table*}[htbp]
\centering
\scriptsize
\setlength{\tabcolsep}{4pt}
\begin{tabular}{@{}l@{\hspace{2pt}}c@{\hspace{2pt}}c@{\hspace{2pt}}c@{\hspace{2pt}}c@{\hspace{2pt}}c@{\hspace{2pt}}c@{}}
\toprule
& \multicolumn{2}{c}{D vs R} & \multicolumn{2}{c}{R vs U-R} & \multicolumn{2}{c}{D vs U-D} \\
\cmidrule(lr){2-3} \cmidrule(lr){4-5} \cmidrule(lr){6-7}
\textbf{Dimension} & {\color{DemocratBlue} \textbf{D}} & {\color{RepublicanRed} \textbf{R}} & {\color{RepublicanRed} \textbf{R}} & {\color{UnaffiliatedBrown} \textbf{U-R}} & {\color{DemocratBlue} \textbf{D}} & {\color{UnaffiliatedBrown} \textbf{U-D}} \\
\midrule
Social & 12.32 $\pm$ 13.35 & {\color{RepublicanRed}12.93 $\pm$ 13.80 (R)$^{\dagger}$} & {\color{RepublicanRed}12.93 $\pm$ 13.80 (R)$^{\dagger}$} & 11.85 $\pm$ 12.83 & {\color{DemocratBlue}12.32 $\pm$ 13.35 (D)$^{\dagger}$} & 11.59 $\pm$ 12.78 \\
Male & 1.24 $\pm$ 4.66 & {\color{RepublicanRed}1.38 $\pm$ 4.91 (R)$^{***}$} & {\color{RepublicanRed}1.38 $\pm$ 4.91 (R)$^{\dagger}$} & 1.16 $\pm$ 4.48 & {\color{DemocratBlue}1.24 $\pm$ 4.66 (D)$^{**}$} & 1.17 $\pm$ 4.50 \\
Female & 1.24 $\pm$ 4.61 & {\color{RepublicanRed}1.44 $\pm$ 4.92 (R)$^{\dagger}$} & {\color{RepublicanRed}1.44 $\pm$ 4.92 (R)$^{\dagger}$} & 1.13 $\pm$ 4.16 & {\color{DemocratBlue}1.24 $\pm$ 4.61 (D)$^{***}$} & 1.15 $\pm$ 4.26 \\
Cog Process & {\color{DemocratBlue}14.57 $\pm$ 13.52 (D)$^{***}$} & 14.26 $\pm$ 13.59 & 14.26 $\pm$ 13.59 & {\color{UnaffiliatedBrown}14.69 $\pm$ 13.23 (U)$^{***}$} & 14.57 $\pm$ 13.52 & {\color{UnaffiliatedBrown}14.64 $\pm$ 13.20 (U)$^{*}$} \\
Pos Emotion & 5.76 $\pm$ 10.07 & 5.79 $\pm$ 10.47 & 5.79 $\pm$ 10.47 & {\color{UnaffiliatedBrown}6.40 $\pm$ 10.81 (U)$^{*}$} & 5.76 $\pm$ 10.07 & {\color{UnaffiliatedBrown}6.33 $\pm$ 10.74 (U)$^{\dagger}$} \\
Neg Emotion & {\color{DemocratBlue}5.41 $\pm$ 9.19 (D)$^{*}$} & 5.36 $\pm$ 9.19 & {\color{RepublicanRed}5.36 $\pm$ 9.19 (R)$^{**}$} & 5.23 $\pm$ 8.67 & {\color{DemocratBlue}5.41 $\pm$ 9.19 (D)$^{***}$} & 5.15 $\pm$ 8.60 \\
Anxiety & 0.82 $\pm$ 3.22 & 0.82 $\pm$ 3.26 & 0.82 $\pm$ 3.26 & {\color{UnaffiliatedBrown}0.90 $\pm$ 3.15 (U)$^{***}$} & 0.82 $\pm$ 3.22 & {\color{UnaffiliatedBrown}0.89 $\pm$ 3.08 (U)$^{***}$} \\
Anger & {\color{DemocratBlue}1.98 $\pm$ 5.95 (D)$^{**}$} & 1.91 $\pm$ 5.88 & {\color{RepublicanRed}1.91 $\pm$ 5.88 (R)$^{\dagger}$} & 1.59 $\pm$ 5.13 & {\color{DemocratBlue}1.98 $\pm$ 5.95 (D)$^{\dagger}$} & 1.63 $\pm$ 5.20 \\
Feel & 1.25 $\pm$ 4.05 & 1.24 $\pm$ 4.08 & 1.24 $\pm$ 4.08 & {\color{UnaffiliatedBrown}1.53 $\pm$ 4.29 (U)$^{\dagger}$} & 1.25 $\pm$ 4.05 & {\color{UnaffiliatedBrown}1.53 $\pm$ 4.34 (U)$^{\dagger}$} \\
Bio & {\color{DemocratBlue}4.60 $\pm$ 8.56 (D)$^{**}$} & 4.52 $\pm$ 8.62 & {\color{RepublicanRed}4.52 $\pm$ 8.62 (R)$^{**}$} & 4.39 $\pm$ 7.94 & {\color{DemocratBlue}4.60 $\pm$ 8.56 (D)$^{***}$} & 4.34 $\pm$ 8.03 \\
Body & {\color{DemocratBlue}1.23 $\pm$ 4.33 (D)$^{**}$} & 1.20 $\pm$ 4.32 & {\color{RepublicanRed}1.20 $\pm$ 4.32 (R)$^{***}$} & 1.04 $\pm$ 3.74 & {\color{DemocratBlue}1.23 $\pm$ 4.33 (D)$^{***}$} & 1.10 $\pm$ 3.97 \\
Health & 1.88 $\pm$ 5.26 & 1.90 $\pm$ 5.38 & 1.90 $\pm$ 5.38 & {\color{UnaffiliatedBrown}2.05 $\pm$ 5.20 (U)$^{***}$} & 1.88 $\pm$ 5.26 & 1.89 $\pm$ 5.04 \\
Affiliation & 2.18 $\pm$ 5.91 & {\color{RepublicanRed}2.33 $\pm$ 6.36 (R)$^{***}$} & 2.33 $\pm$ 6.36 & {\color{UnaffiliatedBrown}2.36 $\pm$ 6.13 (U)$^{*}$} & 2.18 $\pm$ 5.91 & {\color{UnaffiliatedBrown}2.30 $\pm$ 6.08 (U)$^{***}$} \\
Work & 3.03 $\pm$ 6.80 & {\color{RepublicanRed}3.06 $\pm$ 7.02 (R)$^{*}$} & {\color{RepublicanRed}3.06 $\pm$ 7.02 (R)$^{\dagger}$} & 2.62 $\pm$ 6.14 & {\color{DemocratBlue}3.03 $\pm$ 6.80 (D)$^{\dagger}$} & 2.51 $\pm$ 6.03 \\
Leisure & 1.31 $\pm$ 4.78 & {\color{RepublicanRed}1.35 $\pm$ 4.81 (R)$^{**}$} & {\color{RepublicanRed}1.35 $\pm$ 4.81 (R)$^{**}$} & 1.28 $\pm$ 4.65 & {\color{DemocratBlue}1.31 $\pm$ 4.78 (D)$^{*}$} & 1.28 $\pm$ 4.68 \\
Home & 0.44 $\pm$ 2.42 & {\color{RepublicanRed}0.47 $\pm$ 2.53 (R)$^{**}$} & {\color{RepublicanRed}0.47 $\pm$ 2.53 (R)$^{***}$} & 0.42 $\pm$ 2.25 & {\color{DemocratBlue}0.44 $\pm$ 2.42 (D)$^{**}$} & 0.41 $\pm$ 2.20 \\
Money & 0.89 $\pm$ 3.72 & {\color{RepublicanRed}0.95 $\pm$ 3.94 (R)$^{***}$} & {\color{RepublicanRed}0.95 $\pm$ 3.94 (R)$^{\dagger}$} & 0.65 $\pm$ 2.98 & {\color{DemocratBlue}0.89 $\pm$ 3.72 (D)$^{\dagger}$} & 0.64 $\pm$ 3.03 \\
Religion & 0.30 $\pm$ 2.57 & {\color{RepublicanRed}0.35 $\pm$ 2.93 (R)$^{***}$} & {\color{RepublicanRed}0.35 $\pm$ 2.93 (R)$^{***}$} & 0.25 $\pm$ 2.22 & {\color{DemocratBlue}0.30 $\pm$ 2.57 (D)$^{***}$} & 0.25 $\pm$ 2.23 \\
Death & 0.26 $\pm$ 1.88 & {\color{RepublicanRed}0.28 $\pm$ 2.02 (R)$^{**}$} & 0.28 $\pm$ 2.02 & 0.27 $\pm$ 1.93 & 0.26 $\pm$ 1.88 & 0.26 $\pm$ 1.85 \\
\bottomrule
\end{tabular}
\caption{Psycholinguistic Analysis, Expressive Differences. Values shown are mean percentages $\pm$ SD. Letters in parentheses and colored text indicate group with higher values. Columns show three comparisons: {\color{DemocratBlue}Democrat} vs {\color{RepublicanRed}Republican} users (D vs R), {\color{RepublicanRed}Republican} vs {\color{UnaffiliatedBrown}Unaffiliated users matched with Republican users} (R vs U-R), and {\color{DemocratBlue}Democrat} vs {\color{UnaffiliatedBrown}Unaffiliated users matched with Democrat users} (D vs U-D). Significance: $^{*}p < .05$; $^{**}p < 0.00001$; $^{***}p < 10^{-20}$; $^{\dagger}p < 10^{-100}$. Exact p-values available in Supplement F. Significant results highlighted by group color. Both {\color{RepublicanRed}Republican} and {\color{DemocratBlue}Democrat} users tend to use online mental health support forums to describe their day-to-day life events and social connections when compared to {\color{UnaffiliatedBrown}Unaffiliated} users. However, when compared directly, {\color{RepublicanRed}Republicans} appear to use these platforms more for direct reporting of experiences, though {\color{DemocratBlue}Democrats} show higher use of clinical and biological terminology in expressing their distress.}
\label{tab:psycholinguistic_liwc_differences}
\vspace{-15pt}
\end{table*}
}

\newcommand{\clinicaldifferencesliwc}{
\begin{table*}[htbp]
\centering
\label{tab:psycholinguistic-differences-clinical}
\scriptsize
\setlength{\tabcolsep}{4pt}
\begin{tabular}{@{}l@{\hspace{2pt}}c@{\hspace{2pt}}c@{\hspace{2pt}}c@{\hspace{2pt}}c@{\hspace{2pt}}c@{\hspace{2pt}}c@{}}
\toprule
& \multicolumn{2}{c}{D vs R} & \multicolumn{2}{c}{R vs U-R} & \multicolumn{2}{c}{D vs U-D} \\
\cmidrule(lr){2-3} \cmidrule(lr){4-5} \cmidrule(lr){6-7}
\textbf{Dimension} & {\color{DemocratBlue} \textbf{D}} & {\color{RepublicanRed} \textbf{R}} & {\color{RepublicanRed} \textbf{R}} & {\color{UnaffiliatedBrown} \textbf{U-R}} & {\color{DemocratBlue} \textbf{D}} & {\color{UnaffiliatedBrown} \textbf{U-D}} \\
\midrule
Social & 11.47 $\pm$ 8.99 & {\color{RepublicanRed}11.84 $\pm$ 9.33 (R)$^{***}$} & {\color{RepublicanRed}11.96 $\pm$ 9.30 (R)$^{***}$} & 11.27 $\pm$ 8.68 & {\color{DemocratBlue}11.47 $\pm$ 8.98 (D)$^{***}$} & 10.98 $\pm$ 8.55 \\
Male & 0.89 $\pm$ 2.68 & {\color{RepublicanRed}0.99 $\pm$ 2.81 (R)$^{**}$} & {\color{RepublicanRed}1.00 $\pm$ 2.82 (R)$^{**}$} & 0.91 $\pm$ 2.59 & 0.89 $\pm$ 2.68 & 0.87 $\pm$ 2.52 \\
Female & 0.98 $\pm$ 3.05 & {\color{RepublicanRed}1.15 $\pm$ 3.32 (R)$^{***}$} & {\color{RepublicanRed}1.16 $\pm$ 3.32 (R)$^{***}$} & 1.00 $\pm$ 2.96 & 0.98 $\pm$ 3.04 & 1.00 $\pm$ 2.99 \\
Cog Process & {\color{DemocratBlue}16.49 $\pm$ 9.05 (D)$^{***}$} & 15.89 $\pm$ 8.96 & 16.03 $\pm$ 8.89 & {\color{UnaffiliatedBrown}16.53 $\pm$ 8.69 (U)$^{***}$} & 16.52 $\pm$ 9.04 & 16.47 $\pm$ 8.52 \\
Pos Emotion & {\color{DemocratBlue}4.85 $\pm$ 4.92 (D)$^{**}$} & 4.71 $\pm$ 4.89 & 4.76 $\pm$ 4.92 & {\color{UnaffiliatedBrown}4.95 $\pm$ 4.86 (U)$^{**}$} & 4.84 $\pm$ 4.90 & 4.84 $\pm$ 4.77 \\
Neg Emotion & 7.00 $\pm$ 6.98 & {\color{RepublicanRed}7.24 $\pm$ 7.09 (R)$^{**}$} & {\color{RepublicanRed}7.31 $\pm$ 7.09 (R)$^{***}$} & 6.98 $\pm$ 6.74 & 6.98 $\pm$ 6.96 & 6.97 $\pm$ 6.55 \\
Anxiety & 1.78 $\pm$ 3.74 & {\color{RepublicanRed}1.91 $\pm$ 4.01 (R)$^{**}$} & {\color{RepublicanRed}1.95 $\pm$ 4.06 (R)$^{**}$} & 1.86 $\pm$ 3.58 & 1.77 $\pm$ 3.73 & {\color{UnaffiliatedBrown}1.84 $\pm$ 3.48 (U)$^{**}$} \\
Anger & 1.63 $\pm$ 3.34 & 1.63 $\pm$ 3.33 & {\color{RepublicanRed}1.63 $\pm$ 3.30 (R)$^{***}$} & 1.47 $\pm$ 3.06 & {\color{DemocratBlue}1.62 $\pm$ 3.31 (D)$^{**}$} & 1.53 $\pm$ 3.00 \\
Feel & 1.39 $\pm$ 2.58 & 1.41 $\pm$ 2.61 & 1.44 $\pm$ 2.62 & {\color{UnaffiliatedBrown}1.68 $\pm$ 2.75 (U)$^{***}$} & 1.39 $\pm$ 2.57 & {\color{UnaffiliatedBrown}1.65 $\pm$ 2.61 (U)$^{***}$} \\
Bio & 6.19 $\pm$ 6.99 & 6.18 $\pm$ 7.08 & {\color{RepublicanRed}6.26 $\pm$ 7.11 (R)$^{**}$} & 6.01 $\pm$ 6.62 & {\color{DemocratBlue}6.20 $\pm$ 6.98 (D)$^{***}$} & 5.89 $\pm$ 6.56 \\
Body & 1.36 $\pm$ 3.18 & {\color{RepublicanRed}1.43 $\pm$ 3.33 (R)$^{**}$} & {\color{RepublicanRed}1.45 $\pm$ 3.34 (R)$^{***}$} & 1.30 $\pm$ 3.02 & {\color{DemocratBlue}1.36 $\pm$ 3.17 (D)$^{*}$} & 1.32 $\pm$ 3.09 \\
Health & {\color{DemocratBlue}3.85 $\pm$ 5.82 (D)$^{*}$} & 3.77 $\pm$ 5.87 & 3.82 $\pm$ 5.91 & 3.80 $\pm$ 5.57 & {\color{DemocratBlue}3.86 $\pm$ 5.82 (D)$^{**}$} & 3.66 $\pm$ 5.41 \\
Affiliation & 1.90 $\pm$ 3.26 & {\color{RepublicanRed}1.95 $\pm$ 3.30 (R)$^{*}$} & 1.96 $\pm$ 3.31 & 1.99 $\pm$ 3.19 & 1.90 $\pm$ 3.26 & 1.93 $\pm$ 3.14 \\
Work & {\color{DemocratBlue}3.30 $\pm$ 4.61 (D)$^{**}$} & 3.13 $\pm$ 4.55 & {\color{RepublicanRed}3.16 $\pm$ 4.57 (R)$^{***}$} & 2.93 $\pm$ 4.22 & {\color{DemocratBlue}3.30 $\pm$ 4.61 (D)$^{\dagger}$} & 2.84 $\pm$ 4.09 \\
Leisure & 1.04 $\pm$ 2.51 & {\color{RepublicanRed}1.07 $\pm$ 2.55 (R)$^{*}$} & {\color{RepublicanRed}1.09 $\pm$ 2.55 (R)$^{**}$} & 1.03 $\pm$ 2.43 & {\color{DemocratBlue}1.04 $\pm$ 2.50 (D)$^{*}$} & 1.00 $\pm$ 2.39 \\
Home & 0.40 $\pm$ 1.41 & {\color{RepublicanRed}0.43 $\pm$ 1.48 (R)$^{**}$} & {\color{RepublicanRed}0.43 $\pm$ 1.49 (R)$^{*}$} & 0.41 $\pm$ 1.36 & 0.39 $\pm$ 1.40 & 0.40 $\pm$ 1.34 \\
Money & 0.65 $\pm$ 2.06 & 0.66 $\pm$ 2.11 & {\color{RepublicanRed}0.66 $\pm$ 2.12 (R)$^{***}$} & 0.51 $\pm$ 1.71 & {\color{DemocratBlue}0.64 $\pm$ 2.05 (D)$^{***}$} & 0.51 $\pm$ 1.68 \\
Religion & 0.19 $\pm$ 1.07 & {\color{RepublicanRed}0.22 $\pm$ 1.17 (R)$^{**}$} & {\color{RepublicanRed}0.23 $\pm$ 1.19 (R)$^{***}$} & 0.18 $\pm$ 0.96 & {\color{DemocratBlue}0.19 $\pm$ 1.05 (D)$^{**}$} & 0.17 $\pm$ 0.97 \\
Death & 0.41 $\pm$ 1.95 & {\color{RepublicanRed}0.47 $\pm$ 2.02 (R)$^{**}$} & 0.47 $\pm$ 2.01 & 0.46 $\pm$ 1.94 & 0.41 $\pm$ 1.95 & {\color{UnaffiliatedBrown}0.43 $\pm$ 1.85 (U)$^{*}$} \\
\bottomrule
\end{tabular}
\caption{Psycholinguistic Analysis, Clinical Language Differences. Values shown are mean percentages $\pm$ SD. Letters in parentheses and colored text indicate group with higher values. Columns show three comparisons: {\color{DemocratBlue}Democrat} vs {\color{RepublicanRed}Republican} users (D vs R), {\color{RepublicanRed}Republican} vs {\color{UnaffiliatedBrown}Unaffiliated users matched with Republican users} (R vs U-R), and {\color{DemocratBlue}Democrat} vs {\color{UnaffiliatedBrown}Unaffiliated users matched with Democrat users} (D vs U-D). Significance: $^{*}p < .05$; $^{**}p < 0.00001$; $^{***}p < 10^{-20}$; $^{\dagger}p < 10^{-100}$. Exact p-values available in Supplement F. Comparing {\color{RepublicanRed}Republican} and {\color{DemocratBlue}Democrat} users, in clinical posts, we find higher use of biological language among {\color{RepublicanRed}Republican} users, and higher use of health language from {\color{DemocratBlue}Democrat} users, a shift from our expressive analysis of all posts. This is suggestive of a more somatic framing of distress from {\color{RepublicanRed}Republican} users. Compared to {\color{UnaffiliatedBrown}Unaffiliated} users, both {\color{RepublicanRed}Republican} and {\color{DemocratBlue}Democrat} users have more language suggesting discussion of social relationships and day-to-day life alongside clinical language.}
\label{tab:clinical_liwc_differences}
\vspace{-15pt}
\end{table*}
}

\newcommand{\polarizationdifferencesliwc}{
\begin{table*}[htbp]
\centering
\scriptsize
\setlength{\tabcolsep}{4pt}
\begin{tabular}{@{}l@{\hspace{2pt}}c@{\hspace{2pt}}c@{\hspace{2pt}}c@{\hspace{2pt}}c@{\hspace{2pt}}c@{\hspace{2pt}}c@{}}
\toprule
& \multicolumn{2}{c}{D vs R} & \multicolumn{2}{c}{R vs U-R} & \multicolumn{2}{c}{D vs U-D} \\
\cmidrule(lr){2-3} \cmidrule(lr){4-5} \cmidrule(lr){6-7}
\textbf{Dimension} & {\color{DemocratBlue} \textbf{D}} & {\color{RepublicanRed} \textbf{R}} & {\color{RepublicanRed} \textbf{R}} & {\color{UnaffiliatedBrown} \textbf{U-R}} & {\color{DemocratBlue} \textbf{D}} & {\color{UnaffiliatedBrown} \textbf{U-D}} \\
\midrule
Social & 14.92 $\pm$ 11.01 & {\color{RepublicanRed}15.43 $\pm$ 11.29 (R)$^{***}$} & {\color{RepublicanRed}15.44 $\pm$ 11.28 (R)$^{\dagger}$} & 14.43 $\pm$ 10.45 & {\color{DemocratBlue}14.92 $\pm$ 11.00 (D)$^{***}$} & 14.16 $\pm$ 10.46 \\
Male & 1.22 $\pm$ 3.41 & {\color{RepublicanRed}1.34 $\pm$ 3.55 (R)$^{***}$} & {\color{RepublicanRed}1.34 $\pm$ 3.54 (R)$^{***}$} & 1.19 $\pm$ 3.21 & {\color{DemocratBlue}1.22 $\pm$ 3.40 (D)$^{*}$} & 1.18 $\pm$ 3.21 \\
Female & 1.36 $\pm$ 3.94 & {\color{RepublicanRed}1.57 $\pm$ 4.23 (R)$^{***}$} & {\color{RepublicanRed}1.57 $\pm$ 4.22 (R)$^{***}$} & 1.32 $\pm$ 3.68 & 1.36 $\pm$ 3.93 & 1.36 $\pm$ 3.78 \\
Cog Process & {\color{DemocratBlue}15.84 $\pm$ 10.03 (D)$^{*}$} & 15.60 $\pm$ 9.99 & 15.61 $\pm$ 9.99 & {\color{UnaffiliatedBrown}16.34 $\pm$ 9.63 (U)$^{\dagger}$} & 15.86 $\pm$ 10.03 & {\color{UnaffiliatedBrown}16.40 $\pm$ 9.59 (U)$^{***}$} \\
Pos Emotion & {\color{DemocratBlue}5.15 $\pm$ 5.82 (D)$^{*}$} & 5.10 $\pm$ 5.88 & 5.11 $\pm$ 5.88 & {\color{UnaffiliatedBrown}5.36 $\pm$ 5.73 (U)$^{***}$} & 5.15 $\pm$ 5.81 & {\color{UnaffiliatedBrown}5.30 $\pm$ 5.70 (U)$^{**}$} \\
Neg Emotion & 7.76 $\pm$ 9.02 & 7.79 $\pm$ 9.02 & {\color{RepublicanRed}7.77 $\pm$ 8.98 (R)$^{***}$} & 7.45 $\pm$ 8.30 & {\color{DemocratBlue}7.74 $\pm$ 9.00 (D)$^{***}$} & 7.40 $\pm$ 8.12 \\
Anxiety & 0.99 $\pm$ 2.64 & {\color{RepublicanRed}1.02 $\pm$ 2.71 (R)$^{*}$} & {\color{RepublicanRed}1.02 $\pm$ 2.71 (R)$^{***}$} & 1.15 $\pm$ 2.59 & 0.99 $\pm$ 2.63 & {\color{UnaffiliatedBrown}1.15 $\pm$ 2.56 (U)$^{***}$} \\
Anger & {\color{DemocratBlue}3.53 $\pm$ 7.17 (D)$^{*}$} & 3.48 $\pm$ 7.11 & {\color{RepublicanRed}3.46 $\pm$ 7.05 (R)$^{\dagger}$} & 2.88 $\pm$ 6.23 & {\color{DemocratBlue}3.52 $\pm$ 7.14 (D)$^{\dagger}$} & 2.94 $\pm$ 6.27 \\
Feel & 1.32 $\pm$ 2.85 & {\color{RepublicanRed}1.34 $\pm$ 2.94 (R)$^{*}$} & 1.34 $\pm$ 2.90 & {\color{UnaffiliatedBrown}1.66 $\pm$ 3.01 (U)$^{\dagger}$} & 1.32 $\pm$ 2.85 & {\color{UnaffiliatedBrown}1.68 $\pm$ 3.06 (U)$^{\dagger}$} \\
Bio & {\color{DemocratBlue}5.20 $\pm$ 7.32 (D)$^{**}$} & 5.05 $\pm$ 7.18 & {\color{RepublicanRed}5.04 $\pm$ 7.15 (R)$^{***}$} & 4.81 $\pm$ 6.57 & {\color{DemocratBlue}5.19 $\pm$ 7.29 (D)$^{***}$} & 4.84 $\pm$ 6.61 \\
Body & {\color{DemocratBlue}1.52 $\pm$ 3.93 (D)$^{*}$} & 1.48 $\pm$ 3.85 & {\color{RepublicanRed}1.47 $\pm$ 3.83 (R)$^{***}$} & 1.28 $\pm$ 3.36 & {\color{DemocratBlue}1.51 $\pm$ 3.91 (D)$^{***}$} & 1.36 $\pm$ 3.61 \\
Health & 1.85 $\pm$ 3.56 & 1.85 $\pm$ 3.56 & 1.85 $\pm$ 3.56 & {\color{UnaffiliatedBrown}2.03 $\pm$ 3.58 (U)$^{***}$} & 1.85 $\pm$ 3.56 & {\color{UnaffiliatedBrown}1.95 $\pm$ 3.44 (U)$^{**}$} \\
Affiliation & 2.20 $\pm$ 3.90 & {\color{RepublicanRed}2.29 $\pm$ 4.02 (R)$^{**}$} & 2.29 $\pm$ 4.02 & {\color{UnaffiliatedBrown}2.34 $\pm$ 3.85 (U)$^{*}$} & 2.21 $\pm$ 3.90 & {\color{UnaffiliatedBrown}2.29 $\pm$ 3.81 (U)$^{**}$} \\
Work & 2.97 $\pm$ 4.85 & 2.96 $\pm$ 4.90 & {\color{RepublicanRed}2.97 $\pm$ 4.90 (R)$^{\dagger}$} & 2.56 $\pm$ 4.24 & {\color{DemocratBlue}2.98 $\pm$ 4.85 (D)$^{\dagger}$} & 2.51 $\pm$ 4.20 \\
Leisure & 1.13 $\pm$ 2.89 & {\color{RepublicanRed}1.18 $\pm$ 2.98 (R)$^{**}$} & {\color{RepublicanRed}1.18 $\pm$ 2.98 (R)$^{**}$} & 1.10 $\pm$ 2.76 & {\color{DemocratBlue}1.13 $\pm$ 2.89 (D)$^{*}$} & 1.10 $\pm$ 2.72 \\
Home & 0.42 $\pm$ 1.65 & {\color{RepublicanRed}0.45 $\pm$ 1.71 (R)$^{**}$} & {\color{RepublicanRed}0.45 $\pm$ 1.71 (R)$^{**}$} & 0.41 $\pm$ 1.49 & 0.43 $\pm$ 1.65 & 0.42 $\pm$ 1.52 \\
Money & 0.85 $\pm$ 2.70 & {\color{RepublicanRed}0.89 $\pm$ 2.72 (R)$^{**}$} & {\color{RepublicanRed}0.89 $\pm$ 2.73 (R)$^{\dagger}$} & 0.64 $\pm$ 2.14 & {\color{DemocratBlue}0.85 $\pm$ 2.70 (D)$^{\dagger}$} & 0.63 $\pm$ 2.12 \\
Religion & 0.32 $\pm$ 1.84 & {\color{RepublicanRed}0.35 $\pm$ 1.91 (R)$^{**}$} & {\color{RepublicanRed}0.35 $\pm$ 1.90 (R)$^{***}$} & 0.26 $\pm$ 1.52 & {\color{DemocratBlue}0.32 $\pm$ 1.84 (D)$^{***}$} & 0.26 $\pm$ 1.55 \\
Death & 0.39 $\pm$ 1.94 & {\color{RepublicanRed}0.43 $\pm$ 2.09 (R)$^{**}$} & {\color{RepublicanRed}0.43 $\pm$ 2.09 (R)$^{*}$} & 0.42 $\pm$ 1.91 & 0.39 $\pm$ 1.94 & 0.40 $\pm$ 1.86 \\
\bottomrule
\end{tabular}
\caption{Psycholinguistic Analysis, Polarization Language Differences. Values shown are mean percentages $\pm$ SD. Letters in parentheses and colored text indicate group with higher values. Columns show three comparisons: {\color{DemocratBlue}Democrat} vs {\color{RepublicanRed}Republican} users (D vs R), {\color{RepublicanRed}Republican} vs {\color{UnaffiliatedBrown}Unaffiliated users matched with Republican users} (R vs U-R), and {\color{DemocratBlue}Democrat} vs {\color{UnaffiliatedBrown}Unaffiliated users matched with Democrat users} (D vs U-D).  Significance: $^{*}p < .05$; $^{**}p < 0.00001$; $^{***}p < 10^{-20}$; $^{\dagger}p < 10^{-100}$. Exact p-values available in Supplement F. Comparing {\color{RepublicanRed}Republican} and {\color{DemocratBlue}Democrat} users, we find similar patterns to our expressive and clinical analyses, with more cognitive processing language from {\color{DemocratBlue}Democrat} users, and language indicative of social relationships and day‐to‐day life from {\color{RepublicanRed}Republican} users. Compared to {\color{UnaffiliatedBrown}Unaffiliated} users, both {\color{RepublicanRed}Republican} and {\color{DemocratBlue}Democrat} users use less affiliation language in posts with polarization language.}
\label{tab:polarization_liwc_differences}
\vspace{-15pt}
\end{table*}
}

\newcommand{\expressivedifferencessage}{
\begin{table}[ht]
    \centering
    \scriptsize
    \begin{adjustbox}{max width=\textwidth}
    \begin{tabular}{ll}
    \toprule
    \textbf{Label} & \textbf{Words (SAGE Score)} \\
    \midrule
    \textbf{{\color{DemocratBlue} Democrat} and {\color{RepublicanRed} Republican} Users:} & \\
    \textbf{\color{DemocratBlue} Democrats} & \color{DemocratBlue} ADHD (0.6638), diagnosis (0.3582), dose (0.3294), partner (0.3102), brain (0.2934), \\
    & \color{DemocratBlue} shitty (0.2824), psychiatrist (0.2706), system (0.2549), professional (0.2482), \\
    & \color{DemocratBlue} definitely (0.2430), super (0.2298), autistic (0.2261), diagnosed (0.2240), \\
    & \color{DemocratBlue} tend (0.2224), important (0.2221) \\
    \vspace{2pt} \\
    \textbf{\color{RepublicanRed} Republicans} & \color{RepublicanRed} local (0.3421), inside (0.3186), loved (0.3023), please (0.2610), female (0.2467), \\
    & \color{RepublicanRed} degree (0.2242), god (0.2130), business (0.2104), dating (0.2086), dad (0.1945), \\
    & \color{RepublicanRed} date (0.1920), jobs (0.1891), college (0.1846), OCD (0.1820), married (0.1811) \\
    \midrule    
    \textbf{{\color{RepublicanRed} Republican} and {\color{UnaffiliatedBrown} Unaffiliated-R} Users:} & \\
    \textbf{\color{RepublicanRed} Republicans} & \color{RepublicanRed} business (0.6050), local (0.5851), jobs (0.4359), white (0.4209), degree (0.4194), \\
    & \color{RepublicanRed} pay (0.3677), black (0.3647), buy (0.3637), paying (0.3630), married (0.3623), \\
    & \color{RepublicanRed} gay (0.3513), options (0.3455), simply (0.3109), perhaps (0.3039), police (0.2989) \\
    \vspace{2pt} \\
    \textbf{\color{UnaffiliatedBrown} Unaffiliated-R} & \color{UnaffiliatedBrown} BPD (0.9080), ADHD (0.6747), diagnosis (0.5952), trauma (0.5600), partner (0.5554), \\
    & \color{UnaffiliatedBrown} dose (0.5391), relate (0.5180), wanna (0.5176), psychiatrist (0.5046), okay (0.5014), \\
    & \color{UnaffiliatedBrown} definitely (0.4776), struggle (0.4717), hospital (0.4685), myself (0.4672), cry (0.4664) \\
    \midrule
    \textbf{{\color{DemocratBlue} Democrat} and {\color{UnaffiliatedBrown} Unaffiliated-D} Users:} & \\
    \textbf{\color{DemocratBlue} Democrats} & \color{DemocratBlue} certainly (0.4808), bullshit (0.4789), paying (0.4578), white (0.4418), business (0.4333), \\
    & \color{DemocratBlue} society (0.4185), power (0.4007), black (0.3923), gay (0.3890), plenty (0.3753), \\
    & \color{DemocratBlue} history (0.3711), line (0.3675), poor (0.3618), behavior (0.3597), women (0.3574) \\
    \vspace{2pt} \\
    \textbf{\color{UnaffiliatedBrown} Unaffiliated-D} & \color{UnaffiliatedBrown} BPD (0.9890), wanna (0.6503), relate (0.5888), cry (0.5625), struggling (0.5419), \\
    & \color{UnaffiliatedBrown} trauma (0.4898), episode (0.4778), suicidal (0.4635), anxious (0.4603), myself (0.4482), \\
    & \color{UnaffiliatedBrown} tired (0.4423), haha (0.4422), panic (0.4025), feel (0.3914), felt (0.3903) \\
    \bottomrule
    \end{tabular}
    \end{adjustbox}
    \caption{Open Vocabulary-Based Key Phrases, Expressive Differences. Looking first to {\color{DemocratBlue}Democrat} and {\color{RepublicanRed}Republican} users. Democrat users use language that mentions specific mental health disorders, professionals, or treatments. Republican users use language that alludes to social relationships, day-to-day life, and religion. When compared with {\color{UnaffiliatedBrown}Unaffiliated-R} users, {\color{RepublicanRed}Republican} users use language that alludes to identity, occupation and class, and places of particular cultural advocacy. When compared to {\color{UnaffiliatedBrown}Unaffiliated-D} users, {\color{DemocratBlue}Democrat} users use language that hints at a sociocultural and structural approach to mental health, in line with progressive perspectives.}
    \label{tab:expressive_keyphrases}
    \vspace{-15pt}
\end{table}
}

\newcommand{\clinicaldifferencessage}{
\begin{table}[ht]
    \scriptsize
    \centering
    \begin{adjustbox}{max width=\textwidth}
    \begin{tabular}{ll}
    \toprule
    \textbf{Label} & \textbf{Words (SAGE Score)} \\
    \midrule
    \textbf{{\color{DemocratBlue} Democrat} and {\color{RepublicanRed} Republican} Users:} & \\
    \textbf{\color{DemocratBlue} Democrats} & \color{DemocratBlue} \textit{ADHD} (0.6218), partner (0.3039), incredibly (0.2988), \textit{diagnosis} (0.2788), dose (0.2709), \\
    & \color{DemocratBlue} disability (0.2686), function (0.2671), brain (0.2620), psychiatrist (0.2591), \\
    & \color{DemocratBlue} manage (0.2508), important (0.2481), definitely (0.2415), professional (0.2272), \\
    & \color{DemocratBlue} system (0.2194), behavior (0.2164) \\
    \vspace{2pt} \\
    \textbf{\color{RepublicanRed} Republicans} & \color{RepublicanRed} hotline (0.8799), please (0.6842), loved (0.6584), \textit{schizophrenia} (0.5886), \\
    & \color{RepublicanRed} struggling (0.5857), god (0.3313), watch (0.2881), OCD (0.2757), dating (0.2631), \\
    & \color{RepublicanRed} \textit{suicide} (0.2577), woman (0.2541), date (0.2522), college (0.2155), \\
    & \color{RepublicanRed} women (0.2104), \textit{PTSD} (0.2081) \\
    \midrule    
    \textbf{{\color{RepublicanRed} Republican} and {\color{UnaffiliatedBrown} Unaffiliated-R} Users:} & \\
    \textbf{\color{RepublicanRed} Republicans} & \color{RepublicanRed} god (0.4811), married (0.4620), anti (0.4481), men (0.4421), women (0.4178), \\
    & \color{RepublicanRed} man (0.4159), \textit{psychotic} (0.3679), simply (0.3606), drug (0.3526), woman (0.3369), \\
    & \color{RepublicanRed} blood (0.3369), quit (0.3308), drugs (0.3273), poor (0.3241), society (0.3214) \\
    \vspace{2pt} \\
    \textbf{\color{UnaffiliatedBrown} Unaffiliated-R} & \color{UnaffiliatedBrown} BPD (0.6600), struggling (0.5821), partner (0.5537), \textit{ADHD} (0.4857), wanna (0.4757), \\
    & \color{UnaffiliatedBrown} definitely (0.4294), okay (0.4268), crying (0.4001), deserve (0.3873), psychosis (0.3843), \\
    & \color{UnaffiliatedBrown} study (0.3605), \textit{diagnosis} (0.3579), relate (0.3524), sorry (0.3390) \\
    \midrule
    \textbf{{\color{DemocratBlue} Democrat} and {\color{UnaffiliatedBrown} Unaffiliated-D} Users:} & \\
    \textbf{\color{DemocratBlue} Democrats} & \color{DemocratBlue} certainly (0.5907), \textit{condition} (0.4439), number (0.4433), power (0.4296), \\
    & \color{DemocratBlue} drug (0.4250), society (0.4238), disability (0.4128), \textit{conditions} (0.3970), \\
    & \color{DemocratBlue} behavior (0.3953), spectrum (0.3931), line (0.3851), exercise (0.3397), men (0.3384), \\
    & \color{DemocratBlue} ability (0.3209), women (0.2982) \\
    \vspace{2pt} \\
    \textbf{\color{UnaffiliatedBrown} Unaffiliated-D} & \color{UnaffiliatedBrown} BPD (0.8527), cry (0.6145), crying (0.5220), struggling (0.4683), episode (0.4583), \\
    & \color{UnaffiliatedBrown} relate (0.4450), kinda (0.4031), tired (0.3942), episodes (0.3937), myself (0.3889), \\
    & \color{UnaffiliatedBrown} sad (0.3593), \textit{schizophrenia} (0.3569), feel (0.3438), \textit{trauma} (0.3424), \\
    & \color{UnaffiliatedBrown} horrible (0.3385) \\
    \bottomrule
    \end{tabular}
    \end{adjustbox}
    \caption{Open Vocabulary-Based Key Phrases, Clinical Differences. Words from our clinical lexicon are italicized. Looking to posts with clinical language, {\color{RepublicanRed}Republican} users use language that suggests a familiarity with clinical diagnoses, but a reliance on non‐medical sources of care. {\color{DemocratBlue}Democrat} users use language that further emphasizes clinical and medical models of mental illness. When comparing {\color{RepublicanRed}Republican} users with {\color{UnaffiliatedBrown}Unaffiliated-R} users, we observe that god is the most distinctly used word by Republican partisans, echoing Republican partisan culture around religion. Comparing {\color{DemocratBlue}Democrat} users with {\color{UnaffiliatedBrown}Unaffiliated-D} users, we observe Democrat users utilize language suggestive of progressive movements that understand mental illness to be a chronic condition.}
    \label{tab:clinical_keyphrases}
    \vspace{-15pt}
\end{table}
}

\newcommand{\polarizationdifferencessage}{
\begin{table}[ht]
    \scriptsize
    \centering
    \begin{adjustbox}{max width=\textwidth}
    \begin{tabular}{ll}
    \toprule
    \textbf{Label} & \textbf{Words (SAGE Score)} \\
    \midrule
    \textbf{{\color{DemocratBlue} Democrat} and {\color{RepublicanRed} Republican} Users:} & \\
    \textbf{\color{DemocratBlue} Democrats} & \color{DemocratBlue} ADHD (0.6598), diagnosis (0.3523), brain (0.3036), psychiatrist (0.3015), \\
    & \color{DemocratBlue} partner (0.2873), shitty (0.2785), professional (0.2329), certainly (0.2189), \\
    & \color{DemocratBlue} important (0.2149), system (0.2047), diagnosed (0.2007), definitely (0.1996), \\
    & \color{DemocratBlue} trump (0.1992), ADD (0.1982), specific (0.1981) \\
    \vspace{2pt} \\
    \textbf{\color{RepublicanRed} Republicans} & \color{RepublicanRed} ugly (0.3216), degree (0.2359), dad (0.2333), watch (0.2222), \\
    & \color{RepublicanRed} god (0.2196), date (0.2107), heart (0.2091), OCD (0.2085), college (0.2015), \\
    & \color{RepublicanRed} dating (0.2014), married (0.1970), classes (0.1967), baby (0.1952), \\
    & \color{RepublicanRed} \textit{gun} (0.1928), son (0.1818) \\
    \midrule    
    \textbf{{\color{RepublicanRed} Republican} and {\color{UnaffiliatedBrown} Unaffiliated-R} Users:} & \\
    \textbf{\color{RepublicanRed} Republicans} & \color{RepublicanRed} \textit{gun} (0.8192), business (0.5704), white (0.5016), black (0.5016), \\
    & \color{RepublicanRed} jobs (0.4774), married (0.4365), \textit{rape} (0.4092), gay (0.4064), degree (0.4054), \\
    & \color{RepublicanRed} buy (0.4046), police (0.3703), ass (0.3639), perhaps (0.3626), \\
    & \color{RepublicanRed} pay (0.3617), paying (0.3503) \\
    \vspace{2pt} \\
    \textbf{\color{UnaffiliatedBrown} Unaffiliated-R} & \color{UnaffiliatedBrown} BPD (0.8995), struggling (0.7006), ADHD (0.6650), diagnosis (0.5825), \\
    & \color{UnaffiliatedBrown} partner (0.5147), psychiatrist (0.5045), wanna (0.5006), trauma (0.4855), \\
    & \color{UnaffiliatedBrown} relate (0.4607), cry (0.4523), crying (0.4305), suicidal (0.4238), \\
    & \color{UnaffiliatedBrown} myself (0.4206), okay (0.4194), recently (0.3912) \\
    \midrule
    \textbf{{\color{DemocratBlue} Democrat} and {\color{UnaffiliatedBrown} Unaffiliated-D} Users:} & \\
    \textbf{\color{DemocratBlue} Democrats} & \color{DemocratBlue} trump (1.475), certainly (0.4874), \textit{bullshit} (0.4808), white (0.4764), \\
    & \color{DemocratBlue} power (0.4545), business (0.4382), black (0.4356), society (0.4306), \\
    & \color{DemocratBlue} insurance (0.4154), plenty (0.4026), line (0.3973), number (0.3823), \\
    & \color{DemocratBlue} behavior (0.3696), gay (0.3665), police (0.3665) \\
    \vspace{2pt} \\
    \textbf{\color{UnaffiliatedBrown} Unaffiliated-D} & \color{UnaffiliatedBrown} BPD (0.9829), wanna (0.6161), cry (0.5959), struggling (0.5690), \\
    & \color{UnaffiliatedBrown} relate (0.5568), trauma (0.5481), anxious (0.5125), suicidal (0.4753), \\
    & \color{UnaffiliatedBrown} myself (0.4730), tired (0.4496), mood (0.4403), recently (0.4024), \\
    & \color{UnaffiliatedBrown} panic (0.4346), felt (0.4186), feel (0.4057) \\
    \bottomrule
    \end{tabular}
    \end{adjustbox}
    \caption{Open Vocabulary-Based Key Phrases, Polarization Differences. Words from our polarization lexicon are italicized. We find that partisan political rhetoric does appear in expressions of distress from {\color{RepublicanRed}Republican} and {\color{DemocratBlue}Democrat} users. Compared to {\color{UnaffiliatedBrown}Unaffiliated-R} users, partisan cultural positions and debates about identity are reflected in posts from {\color{RepublicanRed}Republican} users. We also observe aspects of structural and economic approaches to mental health reflected in language used distinctly from {\color{DemocratBlue}Democrat} users when compared to {\color{UnaffiliatedBrown}Unaffiliated-D} users.}
    \label{tab:polarization_keyphrases}
\vspace{-15pt}
\end{table}
}

% EXPRESSIVE DIFFERENCES 
\newcommand{\repsdems}{
\begin{table}[htbp]
\centering
\setlength{\tabcolsep}{4pt} % Reduce column spacing
% First tabular (SAGE keywords)
\hspace{-20pt}
\begin{minipage}[t]{0.48\linewidth}
\centering
\tiny
\begin{tabular}{@{}llll@{}}
\toprule
\multicolumn{4}{c}{\textbf{Distinct Keywords, Reps and Dems}} \\ \midrule
{\color{DemocratBlue} \textbf{SAGE}} & {\color{DemocratBlue} \textbf{Democrat}} & {\color{RepublicanRed} \textbf{Republican}} & {\color{RepublicanRed} \textbf{SAGE}} \\
0.6638 & ADHD & local & 0.3421 \\
0.3582 & diagnosis & inside & 0.3186 \\
0.3294 & dose & loved & 0.3023 \\
0.3102 & partner & please & 0.2610 \\
0.2934 & brain & female & 0.2467 \\
0.2824 & shitty & degree & 0.2242 \\
0.2706 & psychiatrist & god & 0.2130 \\
0.2549 & system & business & 0.2104 \\
0.2482 & professional & dating & 0.2086 \\
0.2430 & definitely & dad & 0.1945 \\
0.2298 & super & date & 0.1920 \\
0.2261 & autistic & jobs & 0.1891 \\
0.2240 & diagnosed & college & 0.1846 \\
0.2224 & tend & OCD & 0.1820 \\
0.2221 & important & married & 0.1811 \\ \bottomrule
\bottomrule
\end{tabular}
\vspace{5pt}
\caption{Open vocabulary based key phrases, based on SAGE, {\color{DemocratBlue}Democrat} and {\color{RepublicanRed}Republican} users. Democrat users use language that mentions specific mental health disorders, professionals, or treatments. Republican users use language that alludes to social relationships, day-to-day life, and religion.}
\label{tab:sage_keywords_reps_dems}
\end{minipage}
\hspace{7pt}
% Second tabular (Psycholinguistic Differences)
\begin{minipage}[t]{0.49\linewidth}
\centering
\tiny
\begin{tabular}{@{}llll@{}}
\toprule
\multicolumn{4}{c}{\textbf{Psycholinguistic Differences, Reps and Dems}} \\ \midrule
\textbf{Dimension} & {\color[HTML]{156082} \textbf{Democrat Mean}} & {\color[HTML]{C00000} \textbf{Republican Mean}} & \textbf{Adj. p value} \\
{\color{RepublicanRed}\textbf{Social}} & 12.32\% $\pm$ 13.35\% & 12.93\% $\pm$ 13.80\% & $1.30 \times 10^{-163}$ \\
{\color{RepublicanRed}\textbf{Male}} & 1.24\% $\pm$ 4.66\% & 1.38\% $\pm$ 4.91\% & $1.95 \times 10^{-64}$ \\
{\color{RepublicanRed}\textbf{Female}} & 1.24\% $\pm$ 4.61\% & 1.44\% $\pm$ 4.92\% & $6.37 \times 10^{-131}$ \\
{\color{DemocratBlue}\textbf{Cognitive Processing}} & 14.57\% $\pm$ 13.52\% & 14.26\% $\pm$ 13.59\% & $3.80 \times 10^{-44}$ \\
Positive Emotion & 5.76\% $\pm$ 10.07\% & 5.79\% $\pm$ 10.47\% & 0.131248 \\
{\color{DemocratBlue}\textbf{Negative Emotion}} & 5.41\% $\pm$ 9.19\% & 5.36\% $\pm$ 9.19\% & $3.46 \times 10^{-4}$ \\
Anxiety & 0.82\% $\pm$ 3.22\% & 0.82\% $\pm$ 3.26\% & 0.601142 \\
{\color{DemocratBlue}\textbf{Anger}} & 1.98\% $\pm$ 5.95\% & 1.91\% $\pm$ 5.88\% & $3.65 \times 10^{-12}$ \\
{\color{DemocratBlue}\textbf{Feel}} & 1.25\% $\pm$ 4.05\% & 1.24\% $\pm$ 4.08\% & 0.576841 \\
{\color{DemocratBlue}\textbf{Bio}} & 4.60\% $\pm$ 8.56\% & 4.52\% $\pm$ 8.62\% & $1.43 \times 10^{-8}$ \\
{\color{DemocratBlue}\textbf{Body}} & 1.23\% $\pm$ 4.33\% & 1.20\% $\pm$ 4.32\% & $4.82 \times 10^{-7}$ \\
Health & 1.88\% $\pm$ 5.26\% & 1.90\% $\pm$ 5.38\% & 0.0653815 \\
{\color{RepublicanRed}\textbf{Affiliation}} & 2.18\% $\pm$ 5.91\% & 2.33\% $\pm$ 6.36\% & $1.48 \times 10^{-50}$ \\
{\color{RepublicanRed}\textbf{Work}} & 3.03\% $\pm$ 6.80\% & 3.06\% $\pm$ 7.02\% & $4.21 \times 10^{-3}$ \\
{\color{RepublicanRed}\textbf{Leisure}} & 1.31\% $\pm$ 4.78\% & 1.35\% $\pm$ 4.81\% & $2.84 \times 10^{-5}$ \\
{\color{RepublicanRed}\textbf{Home}} & 0.44\% $\pm$ 2.42\% & 0.47\% $\pm$ 2.53\% & $2.55 \times 10^{-15}$ \\
{\color{RepublicanRed}\textbf{Money}} & 0.89\% $\pm$ 3.72\% & 0.95\% $\pm$ 3.94\% & $4.36 \times 10^{-25}$ \\
{\color{RepublicanRed}\textbf{Religion}} & 0.30\% $\pm$ 2.57\% & 0.35\% $\pm$ 2.93\% & $1.35 \times 10^{-29}$ \\
{\color{RepublicanRed}\textbf{Death}} & 0.26\% $\pm$ 1.88\% & 0.28\% $\pm$ 2.02\% & $2.95 \times 10^{-13}$ \\
\bottomrule
\end{tabular}
\vspace{5pt} 
\caption{Significant LIWC Dimensions, Expressive Differences, {\color{DemocratBlue}Democrat} and {\color{RepublicanRed}Republican} users. Democrat users use more language that suggests processing events that have happened. Republicans use language that suggests describing about day-to-day life events and social relationships.}
\label{tab:liwc_table_reps_dems}
\end{minipage}
\vspace{-18pt}
\end{table}
}

\newcommand{\repsunaffiliated}{
\begin{table}[htbp]
\centering
\setlength{\tabcolsep}{4pt} % Reduce column spacing
% First tabular (SAGE keywords)
\hspace{-20pt}
\begin{minipage}[t]{0.48\linewidth}
\centering
\tiny
\begin{tabular}{@{}llll@{}}
\toprule
\multicolumn{4}{c}{\textbf{Distinct Keywords, Reps and Unaff-R}} \\ \midrule
{\color{RepublicanRed} \textbf{SAGE}} & {\color{RepublicanRed} \textbf{Republican}} & {\color{UnaffiliatedBrown} \textbf{Unaffiliated}} & {\color{UnaffiliatedBrown} \textbf{SAGE}} \\
0.6050 & business & BPD & 0.9080 \\
0.5851 & local & ADHD & 0.6747 \\
0.4359 & jobs & diagnosis & 0.5952 \\
0.4209 & white & trauma & 0.5600 \\
0.4194 & degree & partner & 0.5554 \\
0.3677 & pay & dose & 0.5391 \\
0.3647 & black & relate & 0.5180 \\
0.3637 & buy & wanna & 0.5176 \\
0.3630 & paying & psychiatrist & 0.5046 \\
0.3623 & married & okay & 0.5014 \\
0.3513 & gay & definitely & 0.4766 \\
0.3455 & options & struggle & 0.4717 \\
0.3109 & simply & hospital & 0.4685 \\
0.3039 & perhaps & myself & 0.4672 \\
0.2989 & police & cry & 0.4664
\\ \bottomrule
\bottomrule
\end{tabular}
\vspace{5pt}
\caption{Open vocabulary based key phrases, based on SAGE, {\color{RepublicanRed}Republican} and {\color{UnaffiliatedBrown}Unaffiliated-R} users. Republican users use language that alludes to identity, occupation and class, and places of particular cultural advocacy.}
\label{tab:sage_keywords_rep_unaffiliated}
\end{minipage}
\hspace{7pt}
% Second tabular (Psycholinguistic Differences)
\begin{minipage}[t]{0.49\linewidth}
\centering
\tiny
\begin{tabular}{@{}llll@{}}
\toprule
\multicolumn{4}{c}{\textbf{Psycholinguistic Differences, Reps and Unaff-R}} \\ \midrule
\textbf{Dimension} & {\color{RepublicanRed} \textbf{Republican Mean}} & {\color{UnaffiliatedBrown} \textbf{Unaffiliated Mean}} & \textbf{Adj. p value} \\
{\color{RepublicanRed} \textbf{Social}} & 12.93\% $\pm$ 13.80\% & 11.85\% $\pm$ 12.83\% & $2.23 \times 10^{-308}$ \\
{\color{RepublicanRed} \textbf{Male}} & 1.38\% $\pm$ 4.91\% & 1.16\% $\pm$ 4.48\% & $1.03 \times 10^{-115}$ \\
{\color{RepublicanRed} \textbf{Female}} & 1.44\% $\pm$ 4.92\% & 1.13\% $\pm$ 4.16\% & $1.76 \times 10^{-259}$ \\
{\color{UnaffiliatedBrown} \textbf{Cognitive Processing}} & 14.26\% $\pm$ 13.59\% & 14.69\% $\pm$ 13.23\% & $4.73 \times 10^{-60}$ \\
{\color{UnaffiliatedBrown} \textbf{Positive Emotion}} & 5.79\% $\pm$ 10.47\% & 6.40\% $\pm$ 10.81\% & $< 0.001$ \\
{\color{RepublicanRed} \textbf{Negative Emotion}} & 5.36\% $\pm$ 9.19\% & 5.23\% $\pm$ 8.67\% & $1.77 \times 10^{-13}$ \\
{\color{UnaffiliatedBrown} \textbf{Anxiety}} & 0.82\% $\pm$ 3.26\% & 0.90\% $\pm$ 3.15\% & $7.92 \times 10^{-39}$ \\
{\color{RepublicanRed} \textbf{Anger}} & 1.91\% $\pm$ 5.88\% & 1.59\% $\pm$ 5.13\% & $1.45 \times 10^{-192}$ \\
{\color{UnaffiliatedBrown} \textbf{Feel}} & 1.24\% $\pm$ 4.08\% & 1.53\% $\pm$ 4.29\% & $2.47 \times 10^{-248}$ \\
{\color{RepublicanRed} \textbf{Bio}} & 4.52\% $\pm$ 8.62\% & 4.39\% $\pm$ 7.94\% & $3.18 \times 10^{-16}$ \\
{\color{RepublicanRed} \textbf{Body}} & 1.20\% $\pm$ 4.32\% & 1.04\% $\pm$ 3.74\% & $5.33 \times 10^{-85}$ \\
{\color{UnaffiliatedBrown} \textbf{Health}} & 1.90\% $\pm$ 5.38\% & 2.05\% $\pm$ 5.20\% & $5.12 \times 10^{-46}$ \\
{\color{UnaffiliatedBrown} \textbf{Affiliation}} & 2.33\% $\pm$ 6.36\% & 2.36\% $\pm$ 6.13\% & 0.021 \\
{\color{RepublicanRed} \textbf{Work}} & 3.06\% $\pm$ 7.02\% & 2.62\% $\pm$ 6.14\% & $1.15 \times 10^{-250}$ \\
{\color{RepublicanRed} \textbf{Leisure}} & 1.35\% $\pm$ 4.81\% & 1.28\% $\pm$ 4.65\% & $6.10 \times 10^{-12}$ \\
{\color{RepublicanRed} \textbf{Home}} & 0.47\% $\pm$ 2.53\% & 0.42\% $\pm$ 2.25\% & $3.80 \times 10^{-33}$ \\
{\color{RepublicanRed} \textbf{Money}} & 0.95\% $\pm$ 3.94\% & 0.65\% $\pm$ 2.98\% & $2.23 \times 10^{-308}$ \\
{\color{RepublicanRed} \textbf{Religion}} & 0.35\% $\pm$ 2.93\% & 0.25\% $\pm$ 2.22\% & $1.18 \times 10^{-88}$ \\
Death & 0.28\% $\pm$ 2.02\% & 0.27\% $\pm$ 1.93\% & 0.142 \\
\bottomrule
\end{tabular}
\vspace{5pt} 
\caption{Significant LIWC Dimensions, Expressive Differences, {\color{RepublicanRed}Republican} and {\color{UnaffiliatedBrown}Unaffiliated-R} users. Republican users have lower levels of language suggestive of processing their (health) experiences within online mental health forums, and higher levels of language suggestive of day-to-day issues.}
\label{tab:liwc_table_rep_unaffiliated}
\end{minipage}
\vspace{-18pt}
\end{table}
}

% TO DO 
\newcommand{\demsunaffiliated}{
\begin{table}[htbp]
\centering
\setlength{\tabcolsep}{4pt} % Reduce column spacing
% First tabular (SAGE keywords)
\hspace{-20pt}
\begin{minipage}[t]{0.48\linewidth}
\centering
\tiny
\begin{tabular}{@{}llll@{}}
\toprule
\multicolumn{4}{c}{\textbf{Distinct Keywords, Dems and Unaff-D}} \\ \midrule
{\color{DemocratBlue} \textbf{SAGE}} & {\color{DemocratBlue} \textbf{Democrat}} & {\color{UnaffiliatedBrown} \textbf{Unaffiliated}} & {\color{UnaffiliatedBrown} \textbf{SAGE}} \\
0.4808 & certainly & BPD & 0.9890 \\
0.4789 & bullshit & wanna & 0.6503 \\
0.4578 & paying & relate & 0.5888 \\
0.4418 & white & cry & 0.5625 \\
0.4333 & business & struggling & 0.5419 \\
0.4185 & society & trauma & 0.4898 \\
0.4007 & power & episode & 0.4778 \\
0.3923 & black & suicidal & 0.4635 \\
0.3890 & gay & anxious & 0.4630 \\
0.3753 & plenty & myself & 0.4482 \\
0.3711 & history & tired & 0.4423 \\
0.3675 & line & haha & 0.4422 \\
0.3618 & poor & panic & 0.4025 \\
0.3597 & behavior & feel & 0.3914 \\
0.3574 & women & felt & 0.3903 \\ 
\bottomrule
\bottomrule
\end{tabular}
\vspace{5pt}
\caption{Open vocabulary based key phrases, based on SAGE, {\color{DemocratBlue}Democrat} and {\color{UnaffiliatedBrown}Unaffiliated-D} users. Democrat users use language that hints at a sociocultural and structural approach to mental health, in line with progressive perspectives.}
\label{tab:sage_keywords_dem_unaffiliated}
\end{minipage}
\hspace{7pt}
% Second tabular (Psycholinguistic Differences)
\begin{minipage}[t]{0.49\linewidth}
\centering
\tiny
\begin{tabular}{@{}llll@{}}
\toprule
\multicolumn{4}{c}{\textbf{Psycholinguistic Differences, Dems and Unaff-D}} \\ \midrule
\textbf{Dimension} & {\color{DemocratBlue} \textbf{Democrat Mean}} & {\color{UnaffiliatedBrown} \textbf{Unaffiliated Mean}} & \textbf{Adj. p value} \\
{\color{DemocratBlue} \textbf{Social}} & 12.32\% $\pm$ 13.35\% & 11.59\% $\pm$ 12.78\% & $1.17 \times 10^{-156}$ \\
{\color{DemocratBlue} \textbf{Male}} & 1.24\% $\pm$ 4.66\% & 1.17\% $\pm$ 4.50\% & $5.05 \times 10^{-12}$ \\
{\color{DemocratBlue} \textbf{Female}} & 1.24\% $\pm$ 4.61\% & 1.15\% $\pm$ 4.26\% & $1.46 \times 10^{-22}$ \\
{\color{UnaffiliatedBrown} \textbf{Cognitive Processing}} & 14.57\% $\pm$ 13.52\% & 14.64\% $\pm$ 13.20\% & 0.016 \\
{\color{UnaffiliatedBrown} \textbf{Positive Emotion}} & 5.76\% $\pm$ 10.07\% & 6.33\% $\pm$ 10.74\% & $1.10 \times 10^{-146}$ \\
{\color{DemocratBlue} \textbf{Negative Emotion}} & 5.41\% $\pm$ 9.19\% & 5.15\% $\pm$ 8.60\% & $1.12 \times 10^{-45}$ \\
{\color{UnaffiliatedBrown} \textbf{Anxiety}} & 0.82\% $\pm$ 3.22\% & 0.89\% $\pm$ 3.08\% & $1.45 \times 10^{-22}$ \\
{\color{DemocratBlue} \textbf{Anger}} & 1.98\% $\pm$ 5.95\% & 1.63\% $\pm$ 5.20\% & $3.05 \times 10^{-206}$ \\
{\color{UnaffiliatedBrown} \textbf{Feel}} & 1.25\% $\pm$ 4.05\% & 1.53\% $\pm$ 4.34\% & $2.12 \times 10^{-217}$ \\
{\color{DemocratBlue} \textbf{Bio}} & 4.60\% $\pm$ 8.56\% & 4.34\% $\pm$ 8.03\% & $3.15 \times 10^{-52}$ \\
{\color{DemocratBlue} \textbf{Body}} & 1.23\% $\pm$ 4.33\% & 1.10\% $\pm$ 3.97\% & $2.81 \times 10^{-50}$ \\
Health & 1.88\% $\pm$ 5.26\% & 1.89\% $\pm$ 5.04\% & 0.328 \\
{\color{UnaffiliatedBrown} \textbf{Affiliation}} & 2.18\% $\pm$ 5.91\% & 2.30\% $\pm$ 6.08\% & $2.98 \times 10^{-23}$ \\
{\color{DemocratBlue} \textbf{Work}} & 3.03\% $\pm$ 6.80\% & 2.51\% $\pm$ 6.03\% & $2.23 \times 10^{-308}$ \\
{\color{DemocratBlue} \textbf{Leisure}} & 1.31\% $\pm$ 4.78\% & 1.28\% $\pm$ 4.68\% & $1.42 \times 10^{-3}$ \\
{\color{DemocratBlue} \textbf{Home}} & 0.44\% $\pm$ 2.42\% & 0.41\% $\pm$ 2.20\% & $4.75 \times 10^{-12}$ \\
{\color{DemocratBlue} \textbf{Money}} & 0.89\% $\pm$ 3.72\% & 0.64\% $\pm$ 3.03\% & $8.20 \times 10^{-281}$ \\
{\color{DemocratBlue} \textbf{Religion}} & 0.30\% $\pm$ 2.57\% & 0.25\% $\pm$ 2.23\% & $3.44 \times 10^{-24}$ \\
Death & 0.26\% $\pm$ 1.88\% & 0.26\% $\pm$ 1.85\% & 0.126 \\
\bottomrule
\end{tabular}
\vspace{5pt} 
\caption{Significant LIWC Dimensions, Expressive Differences, {\color{DemocratBlue}Democrat} and {\color{UnaffiliatedBrown}Unaffiliated-D} users. Democrat users use language indicative of conversations around day-to-day life and social relationships, and using health language to a roughly equal extent.}
\label{tab:liwc_table_dem_unaffiliated}
\end{minipage}
\vspace{-18pt}
\end{table}
}

\newcommand{\repsdemsclinical}{
\begin{table}[htbp]
\centering
\setlength{\tabcolsep}{4pt} % Reduce column spacing
% First tabular (SAGE keywords)
\hspace{-20pt}
\begin{minipage}[t]{0.48\linewidth}
\centering
\tiny
\begin{tabular}{@{}llll@{}}
\toprule
\multicolumn{4}{c}{\textbf{Distinct Keywords, Reps and Dems (Clinical)}} \\ \midrule
{\color{DemocratBlue} \textbf{SAGE}} & {\color{DemocratBlue} \textbf{Democrat}} & {\color{RepublicanRed} \textbf{Republican}} & {\color{RepublicanRed} \textbf{SAGE}} \\
0.6218 & \textit{ADHD} & hotline & 0.8799 \\
0.3039 & partner & please & 0.6842 \\
0.2988 & incredibly & loved & 0.6584 \\
0.2788 & \textit{diagnosis} & \textit{schizophrenia} & 0.5886 \\
0.2770 & dose & struggling & 0.5857 \\
0.2686 & disability & god & 0.3313 \\
0.2671 & function & watch & 0.2881 \\
0.2620 & brain & OCD & 0.2757 \\
0.2591 & psychiatrist & dating & 0.2631 \\
0.2508 & manage & \textit{suicide} & 0.2577 \\
0.2481 & important & woman & 0.2541 \\
0.2415 & definitely & date & 0.2522 \\
0.2272 & professional & college & 0.2155 \\
0.2194 & system & women & 0.2104 \\
0.2164 & behavior & \textit{PTSD} & 0.2081
\\ \bottomrule
\bottomrule
\end{tabular}
\vspace{5pt}
\caption{Open vocabulary based key phrases in posts with clinical words, based on SAGE, {\color{DemocratBlue}Democrat} and {\color{RepublicanRed}Republican} users. Words from our clinical lexicon are italicized. Republican users use language that suggests a familiarity with clinical diagnoses, but a reliance on non-medical sources of care. Democrat users use language that further emphasizes clinical and medical models of mental illness.}
\label{tab:sage_keywords_reps_dems_clinical}
\end{minipage}
\hspace{7pt}
% Second tabular (Psycholinguistic Differences)
\begin{minipage}[t]{0.49\linewidth}
\centering
\tiny
\begin{tabular}{@{}llll@{}}
\toprule
\multicolumn{4}{c}{\textbf{Psycholinguistic Differences, Reps and Dems (Clinical)}} \\ \midrule
\textbf{Dimension} & {\color[HTML]{156082} \textbf{Democrat Mean}} & {\color[HTML]{C00000} \textbf{Republican Mean}} & \textbf{Adj. p value} \\
{\color{RepublicanRed} \textbf{Social}} & 11.47\% $\pm$ 8.99\% & 11.84\% $\pm$ 9.33\% & $2.24 \times 10^{-22}$ \\
{\color{RepublicanRed} \textbf{Male}} & 0.89\% $\pm$ 2.68\% & 0.99\% $\pm$ 2.81\% & $5.07 \times 10^{-17}$ \\
{\color{RepublicanRed} \textbf{Female}} & 0.98\% $\pm$ 3.05\% & 1.15\% $\pm$ 3.32\% & $3.19 \times 10^{-39}$ \\
{\color{DemocratBlue} \textbf{Cognitive Processing}} & 16.49\% $\pm$ 9.05\% & 15.89\% $\pm$ 8.96\% & $3.30 \times 10^{-60}$ \\
{\color{DemocratBlue} \textbf{Positive Emotion}} & 4.85\% $\pm$ 4.92\% & 4.71\% $\pm$ 4.89\% & $7.60 \times 10^{-11}$ \\
{\color{RepublicanRed} \textbf{Negative Emotion}} & 7.00\% $\pm$ 6.98\% & 7.24\% $\pm$ 7.09\% & $8.80 \times 10^{-16}$ \\
{\color{RepublicanRed} \textbf{Anxiety}} & 1.78\% $\pm$ 3.74\% & 1.91\% $\pm$ 4.01\% & $3.41 \times 10^{-17}$ \\
{\color{RepublicanRed} \textbf{Anger}} & 1.63\% $\pm$ 3.34\% & 1.63\% $\pm$ 3.33\% & 0.943 \\
Feel & 1.39\% $\pm$ 2.58\% & 1.41\% $\pm$ 2.61\% & 0.050 \\
Bio & 6.19\% $\pm$ 6.99\% & 6.18\% $\pm$ 7.08\% & 0.829 \\
{\color{RepublicanRed} \textbf{Body}} & 1.36\% $\pm$ 3.18\% & 1.43\% $\pm$ 3.33\% & $1.17 \times 10^{-6}$ \\
{\color{DemocratBlue} \textbf{Health}} & 3.85\% $\pm$ 5.82\% & 3.77\% $\pm$ 5.87\% & $4.64 \times 10^{-4}$ \\
{\color{RepublicanRed} \textbf{Affiliation}} & 1.90\% $\pm$ 3.26\% & 1.95\% $\pm$ 3.30\% & $1.04 \times 10^{-3}$ \\
{\color{DemocratBlue} \textbf{Work}} & 3.30\% $\pm$ 4.61\% & 3.13\% $\pm$ 4.55\% & $2.08 \times 10^{-17}$ \\
{\color{RepublicanRed} \textbf{Leisure}} & 1.04\% $\pm$ 2.51\% & 1.07\% $\pm$ 2.55\% & $5.69 \times 10^{-3}$ \\
{\color{RepublicanRed} \textbf{Home}} & 0.40\% $\pm$ 1.41\% & 0.43\% $\pm$ 1.48\% & $5.79 \times 10^{-7}$ \\
Money & 0.65\% $\pm$ 2.06\% & 0.66\% $\pm$ 2.11\% & 0.158 \\
{\color{RepublicanRed} \textbf{Religion}} & 0.19\% $\pm$ 1.07\% & 0.22\% $\pm$ 1.17\% & $1.39 \times 10^{-10}$ \\
{\color{RepublicanRed} \textbf{Death}} & 0.41\% $\pm$ 1.95\% & 0.47\% $\pm$ 2.02\% & $1.20 \times 10^{-11}$ \\
\bottomrule
\end{tabular}
\vspace{5pt} 
\caption{Significant LIWC Dimensions, Clinical Differences, {\color{DemocratBlue}Democrat} and {\color{RepublicanRed}Republican} users. In clinical posts, we find higher use of biological language among Republican users, and higher use of health language from Democrat users, a shift from our analysis of all posts. This is suggestive of a more somatic framing of distress from Republican users.}
\label{tab:liwc_table_reps_dems_clinical}
\end{minipage}
\vspace{-18pt}
\end{table}
}

\newcommand{\repsunaffiliatedclinical}{
\begin{table}[htbp]
\centering
\setlength{\tabcolsep}{4pt} % Reduce column spacing
% First tabular (SAGE keywords)
\hspace{-20pt}
\begin{minipage}[t]{0.48\linewidth}
\centering
\tiny
\begin{tabular}{@{}llll@{}}
\toprule
\multicolumn{4}{c}{\textbf{Distinct Keywords, Reps and Unaff-R (Clinical)}} \\ \midrule
{\color{RepublicanRed} \textbf{SAGE}} & {\color{RepublicanRed} \textbf{Republican}} & {\color{UnaffiliatedBrown} \textbf{Unaffiliated}} & {\color{UnaffiliatedBrown} \textbf{SAGE}} \\
0.4811 & god & BPD & 0.6600 \\
0.4620 & married & struggling & 0.5821 \\
0.4481 & anti & partner & 0.5537 \\
0.4421 & men & \textit{ADHD} & 0.4857 \\
0.4178 & women & wanna & 0.4757 \\
0.4159 & man & definitely & 0.4294 \\
0.3679 & \textit{psychotic} & okay & 0.4268 \\
0.3606 & simply & crying & 0.4163 \\
0.3526 & drug & cry & 0.4001 \\
0.3426 & woman & deserve & 0.3873 \\
0.3369 & blood & psychosis & 0.3834 \\
0.3367 & quit & study & 0.3605 \\
0.3273 & drugs & \textit{diagnosis} & 0.3579 \\
0.3241 & poor & relate & 0.3524 \\
0.3214 & society & sorry & 0.3390 \\ \bottomrule
\bottomrule
\end{tabular}
\vspace{5pt}
\caption{Open vocabulary based key phrases in posts with clinical language, based on SAGE, {\color{RepublicanRed}Republican} and {\color{UnaffiliatedBrown}Unaffiliated-R} users. Words from our clinical lexicon are italicized. Republican partisan culture around religion influences expressions of distress---\republicanword{god} is the most distinctly used word by Republican users.}
\label{tab:sage_keywords_rep_unaffiliated_clinical}
\end{minipage}
\hspace{7pt}
% Second tabular (Psycholinguistic Differences)
\begin{minipage}[t]{0.49\linewidth}
\centering
\tiny
\begin{tabular}{@{}llll@{}}
\toprule
\multicolumn{4}{c}{\textbf{Psycholinguistic Differences, Reps and Unaff-R (Clinical)}} \\ \midrule
\textbf{Dimension} & {\color{RepublicanRed} \textbf{Republican Mean}} & {\color{UnaffiliatedBrown} \textbf{Unaffiliated Mean}} & \textbf{Adj. p value} \\
{\color{RepublicanRed} \textbf{Social}} & 11.96\% $\pm$ 9.30\% & 11.27\% $\pm$ 8.68\% & $1.39 \times 10^{-58}$ \\
{\color{RepublicanRed} \textbf{Male}} & 1.00\% $\pm$ 2.82\% & 0.91\% $\pm$ 2.59\% & $1.04 \times 10^{-13}$ \\
{\color{RepublicanRed} \textbf{Female}} & 1.16\% $\pm$ 3.32\% & 1.00\% $\pm$ 2.96\% & $3.34 \times 10^{-26}$ \\
{\color{UnaffiliatedBrown} \textbf{Cognitive Processing}} & 16.03\% $\pm$ 8.89\% & 16.53\% $\pm$ 8.69\% & $1.09 \times 10^{-32}$ \\
{\color{UnaffiliatedBrown} \textbf{Positive Emotion}} & 4.76\% $\pm$ 4.92\% & 4.95\% $\pm$ 4.86\% & $4.56 \times 10^{-16}$ \\
{\color{RepublicanRed} \textbf{Negative Emotion}} & 7.31\% $\pm$ 7.09\% & 6.98\% $\pm$ 6.74\% & $3.27 \times 10^{-24}$ \\
{\color{RepublicanRed} \textbf{Anxiety}} & 1.95\% $\pm$ 4.06\% & 1.86\% $\pm$ 3.58\% & $1.63 \times 10^{-6}$ \\
{\color{RepublicanRed} \textbf{Anger}} & 1.63\% $\pm$ 3.30\% & 1.47\% $\pm$ 3.06\% & $3.93 \times 10^{-29}$ \\
{\color{UnaffiliatedBrown} \textbf{Feel}} & 1.44\% $\pm$ 2.62\% & 1.68\% $\pm$ 2.75\% & $6.03 \times 10^{-83}$ \\
{\color{RepublicanRed} \textbf{Bio}} & 6.26\% $\pm$ 7.11\% & 6.01\% $\pm$ 6.62\% & $3.97 \times 10^{-14}$ \\
{\color{RepublicanRed} \textbf{Body}} & 1.45\% $\pm$ 3.34\% & 1.30\% $\pm$ 3.02\% & $3.33 \times 10^{-24}$ \\
Health & 3.82\% $\pm$ 5.91\% & 3.80\% $\pm$ 5.57\% & 0.319 \\
Affiliation & 1.96\% $\pm$ 3.31\% & 1.99\% $\pm$ 3.19\% & 0.082 \\
{\color{RepublicanRed} \textbf{Work}} & 3.16\% $\pm$ 4.57\% & 2.93\% $\pm$ 4.22\% & $4.15 \times 10^{-27}$ \\
{\color{RepublicanRed} \textbf{Leisure}} & 1.09\% $\pm$ 2.55\% & 1.03\% $\pm$ 2.43\% & $2.12 \times 10^{-6}$ \\
{\color{RepublicanRed} \textbf{Home}} & 0.43\% $\pm$ 1.49\% & 0.41\% $\pm$ 1.36\% & $1.16 \times 10^{-4}$ \\
{\color{RepublicanRed} \textbf{Money}} & 0.66\% $\pm$ 2.12\% & 0.51\% $\pm$ 1.71\% & $6.57 \times 10^{-68}$ \\
{\color{RepublicanRed} \textbf{Religion}} & 0.23\% $\pm$ 1.19\% & 0.18\% $\pm$ 0.96\% & $2.19 \times 10^{-24}$ \\
Death & 0.47\% $\pm$ 2.01\% & 0.46\% $\pm$ 1.94\% & 0.080 \\\bottomrule
\end{tabular}
\vspace{5pt} 
\caption{Significant LIWC Dimensions, Clinical Differences, {\color{RepublicanRed}Republican} and {\color{UnaffiliatedBrown}Unaffiliated-R} users. Republican users have higher levels of language within their posts centered on social relationships and day to day life, similar to Democrat users.}
\label{tab:liwc_table_rep_unaffiliated_clinical}
\end{minipage}
\vspace{-18pt}
\end{table}
}

\newcommand{\demsunaffiliatedclinical}{
\begin{table}[htbp]
\centering
\setlength{\tabcolsep}{4pt} % Reduce column spacing
% First tabular (SAGE keywords)
\hspace{-20pt}
\begin{minipage}[t]{0.48\linewidth}
\centering
\tiny
\begin{tabular}{@{}llll@{}}
\toprule
\multicolumn{4}{c}{\textbf{Distinct Keywords, Dems and Unaff-D (Clinical)}} \\ \midrule
{\color{DemocratBlue} \textbf{SAGE}} & {\color{DemocratBlue} \textbf{Democrat}} & {\color{UnaffiliatedBrown} \textbf{Unaffiliated}} & {\color{UnaffiliatedBrown} \textbf{SAGE}} \\
0.5907 & certainly & BPD & 0.8527 \\
0.4439 & \textit{condition} & cry & 0.6145 \\
0.4433 & number & crying & 0.5220 \\
0.4296 & power & struggling & 0.4683 \\
0.4250 & drug & episode & 0.4583 \\
0.4238 & society & relate & 0.4450 \\
0.4128 & disability & kinda & 0.4031 \\
0.3970 & \textit{conditions} & tired & 0.3942 \\
0.3953 & behavior & episodes & 0.3937 \\
0.3931 & spectrum & myself & 0.3889 \\
0.3851 & line & sad & 0.3593 \\
0.3397 & exercise & \textit{schizophrenia} & 0.3569 \\
0.3384 & men & feel & 0.3438 \\
0.3209 & ability & \textit{trauma} & 0.3424 \\
0.2982 & women & horrible & 0.3385 \\ \bottomrule
\bottomrule
\end{tabular}
\vspace{5pt}
\caption{Open vocabulary based key phrases from clinical posts, based on SAGE, {\color{DemocratBlue}Democrat} and {\color{UnaffiliatedBrown}Unaffiliated-D} users. Words from our clinical lexicon are italicized. Democrat users utilize language suggestive of progressive movements that understand mental illness to be a chronic condition.}
\label{tab:sage_keywords_dem_unaffiliated_clinical}
\end{minipage}
\hspace{7pt}
% Second tabular (Psycholinguistic Differences)
\begin{minipage}[t]{0.49\linewidth}
\centering
\tiny
\begin{tabular}{@{}llll@{}}
\toprule
\multicolumn{4}{c}{\textbf{Psycholinguistic Differences, Dems and Unaff-D (Clinical)}} \\ \midrule
\textbf{Dimension} & {\color{DemocratBlue} \textbf{Democrat Mean}} & {\color{UnaffiliatedBrown} \textbf{Unaffiliated Mean}} & \textbf{Adj. p value} \\
{\color{DemocratBlue} \textbf{Social}} & 11.47\% $\pm$ 8.98\% & 10.98\% $\pm$ 8.55\% & $6.23 \times 10^{-31}$ \\
Male & 0.89\% $\pm$ 2.68\% & 0.87\% $\pm$ 2.52\% & 0.130 \\
Female & 0.98\% $\pm$ 3.04\% & 1.00\% $\pm$ 2.99\% & 0.116 \\
Cognitive Processing & 16.52\% $\pm$ 9.04\% & 16.47\% $\pm$ 8.52\% & 0.292 \\
Positive Emotion & 4.84\% $\pm$ 4.90\% & 4.84\% $\pm$ 4.77\% & 0.769 \\
Negative Emotion & 6.98\% $\pm$ 6.96\% & 6.97\% $\pm$ 6.55\% & 0.769 \\
{\color{UnaffiliatedBrown} \textbf{Anxiety}} & 1.77\% $\pm$ 3.73\% & 1.84\% $\pm$ 3.48\% & $2.76 \times 10^{-5}$ \\
{\color{DemocratBlue} \textbf{Anger}} & 1.62\% $\pm$ 3.31\% & 1.53\% $\pm$ 3.00\% & $2.70 \times 10^{-10}$ \\
{\color{UnaffiliatedBrown} \textbf{Feel}} & 1.39\% $\pm$ 2.57\% & 1.65\% $\pm$ 2.61\% & $9.87 \times 10^{-92}$ \\
{\color{DemocratBlue} \textbf{Bio}} & 6.20\% $\pm$ 6.98\% & 5.89\% $\pm$ 6.56\% & $3.04 \times 10^{-20}$ \\
{\color{DemocratBlue} \textbf{Body}} & 1.36\% $\pm$ 3.17\% & 1.32\% $\pm$ 3.09\% & 0.014 \\
{\color{DemocratBlue} \textbf{Health}} & 3.86\% $\pm$ 5.82\% & 3.66\% $\pm$ 5.41\% & $7.36 \times 10^{-14}$ \\
Affiliation & 1.90\% $\pm$ 3.26\% & 1.93\% $\pm$ 3.14\% & 0.080 \\
{\color{DemocratBlue} \textbf{Work}} & 3.30\% $\pm$ 4.61\% & 2.84\% $\pm$ 4.09\% & $4.60 \times 10^{-108}$ \\
{\color{DemocratBlue} \textbf{Leisure}} & 1.04\% $\pm$ 2.50\% & 1.00\% $\pm$ 2.39\% & 0.002 \\
Home & 0.39\% $\pm$ 1.40\% & 0.40\% $\pm$ 1.34\% & 0.351 \\
{\color{DemocratBlue} \textbf{Money}} & 0.64\% $\pm$ 2.05\% & 0.51\% $\pm$ 1.68\% & $1.15 \times 10^{-46}$ \\
{\color{DemocratBlue} \textbf{Religion}} & 0.19\% $\pm$ 1.05\% & 0.17\% $\pm$ 0.97\% & $3.53 \times 10^{-5}$ \\
{\color{UnaffiliatedBrown} \textbf{Death}} & 0.41\% $\pm$ 1.95\% & 0.43\% $\pm$ 1.85\% & 0.040 \\
\bottomrule
\end{tabular}
\vspace{5pt} 
\caption{Significant LIWC Dimensions, Clinical Differences, {\color{DemocratBlue}Democrat} and {\color{UnaffiliatedBrown}Unaffiliated-D} users. Democrat users have higher levels of language within their posts centered on social relationships and day to day life.}
\label{tab:liwc_table_dem_unaffiliated_clinical}
\end{minipage}
\vspace{-18pt}
\end{table}
}

\newcommand{\repsdemspolarization}{
\begin{table}[htbp]
\centering
\setlength{\tabcolsep}{4pt} % Reduce column spacing
% First tabular (SAGE keywords)
\hspace{-20pt}
\begin{minipage}[t]{0.48\linewidth}
\centering
\tiny
\begin{tabular}{@{}llll@{}}
\toprule
\multicolumn{4}{c}{\textbf{Distinct Keywords, Reps and Dems (Polarization)}} \\ \midrule
{\color{DemocratBlue} \textbf{SAGE}} & {\color{DemocratBlue} \textbf{Democrat}} & {\color{RepublicanRed} \textbf{Republican}} & {\color{RepublicanRed} \textbf{SAGE}} \\
0.6598 & ADHD & ugly & 0.3216 \\
0.3523 & diagnosis & degree & 0.2359 \\
0.3036 & brain & dad & 0.2333 \\
0.3015 & psychiatrist & watch & 0.2222 \\
0.2873 & partner & god & 0.2196 \\
0.2785 & shitty & date & 0.2107 \\
0.2329 & professional & heart & 0.2091 \\
0.2189 & certainly & OCD & 0.2085 \\
0.2149 & important & college & 0.2015 \\
0.2047 & system & dating & 0.2014 \\
0.2007 & diagnosed & married & 0.1970 \\
0.1996 & definitely & classes & 0.1967 \\
0.1992 & trump & baby & 0.1952 \\
0.1982 & ADD & \textit{gun} & 0.1928 \\
0.1981 & specific & son & 0.1818 \\ \bottomrule
\bottomrule
\end{tabular}
\vspace{5pt}
\caption{Open vocabulary based key phrases in posts with polarization words, based on SAGE, {\color{DemocratBlue}Democrat} and {\color{RepublicanRed}Republican} users. Words from the polarization language lexicon are italicized. Partisan political rhetoric does appear in expressions of distress from both {\color{DemocratBlue}Democrat} and {\color{RepublicanRed}Republican} users.} 
\label{tab:sage_keywords_reps_dems_polarization}
\end{minipage}
\hspace{7pt}
% Second tabular (Psycholinguistic Differences)
\begin{minipage}[t]{0.49\linewidth}
\centering
\tiny
\begin{tabular}{@{}llll@{}}
\toprule
\multicolumn{4}{c}{\textbf{Psycholinguistic Differences, Reps and Dems (Polarization)}} \\ \midrule
\textbf{Dimension} & {\color[HTML]{156082} \textbf{Democrat Mean}} & {\color[HTML]{C00000} \textbf{Republican Mean}} & \textbf{Adj. p value} \\
{\color{RepublicanRed} \textbf{Social}} & 14.92\% $\pm$ 11.01\% & 15.43\% $\pm$ 11.29\% & $1.10 \times 10^{-59}$ \\
{\color{RepublicanRed} \textbf{Male}} & 1.22\% $\pm$ 3.41\% & 1.34\% $\pm$ 3.55\% & $1.10 \times 10^{-34}$ \\
{\color{RepublicanRed} \textbf{Female}} & 1.36\% $\pm$ 3.94\% & 1.57\% $\pm$ 4.23\% & $3.68 \times 10^{-75}$ \\
{\color{DemocratBlue} \textbf{Cognitive Processing}} & 15.84\% $\pm$ 10.03\% & 15.60\% $\pm$ 9.99\% & $6.53 \times 10^{-18}$ \\
{\color{DemocratBlue} \textbf{Positive Emotion}} & 5.15\% $\pm$ 5.82\% & 5.10\% $\pm$ 5.88\% & $9.83 \times 10^{-4}$ \\
Negative Emotion & 7.76\% $\pm$ 9.02\% & 7.79\% $\pm$ 9.02\% & 0.212 \\
{\color{RepublicanRed} \textbf{Anxiety}} & 0.99\% $\pm$ 2.64\% & 1.02\% $\pm$ 2.71\% & $1.77 \times 10^{-4}$ \\
{\color{DemocratBlue} \textbf{Anger}} & 3.53\% $\pm$ 7.17\% & 3.48\% $\pm$ 7.11\% & $5.26 \times 10^{-3}$ \\
{\color{RepublicanRed} \textbf{Feel}} & 1.32\% $\pm$ 2.85\% & 1.34\% $\pm$ 2.94\% & $2.07 \times 10^{-3}$ \\
{\color{DemocratBlue} \textbf{Bio}} & 5.20\% $\pm$ 7.32\% & 5.05\% $\pm$ 7.18\% & $2.70 \times 10^{-13}$ \\
{\color{DemocratBlue} \textbf{Body}} & 1.52\% $\pm$ 3.93\% & 1.48\% $\pm$ 3.85\% & $1.85 \times 10^{-4}$ \\
Health & 1.85\% $\pm$ 3.56\% & 1.85\% $\pm$ 3.56\% & 0.850 \\
{\color{RepublicanRed} \textbf{Affiliation}} & 2.20\% $\pm$ 3.90\% & 2.29\% $\pm$ 4.02\% & $2.63 \times 10^{-14}$ \\
Work & 2.97\% $\pm$ 4.85\% & 2.96\% $\pm$ 4.90\% & 0.601 \\
{\color{RepublicanRed} \textbf{Leisure}} & 1.13\% $\pm$ 2.89\% & 1.18\% $\pm$ 2.98\% & $2.94 \times 10^{-10}$ \\
{\color{RepublicanRed} \textbf{Home}} & 0.42\% $\pm$ 1.65\% & 0.45\% $\pm$ 1.71\% & $1.40 \times 10^{-7}$ \\
{\color{RepublicanRed} \textbf{Money}} & 0.85\% $\pm$ 2.70\% & 0.89\% $\pm$ 2.72\% & $4.51 \times 10^{-5}$ \\
{\color{RepublicanRed} \textbf{Religion}} & 0.32\% $\pm$ 1.84\% & 0.35\% $\pm$ 1.91\% & $7.67 \times 10^{-10}$ \\
{\color{RepublicanRed} \textbf{Death}} & 0.39\% $\pm$ 1.94\% & 0.43\% $\pm$ 2.09\% & $9.07 \times 10^{-13}$ \\
\bottomrule
\end{tabular}
\vspace{5pt} 
\caption{Significant LIWC Dimensions, Polarization Differences, Democrat  and Republican users. We find similar patterns to our expressive and clinical analyses, with more cognitive processing language from Democrat users, and language indicative of social relationships and day-to-day life from Republican users.}
\label{tab:liwc_table_reps_dems_polarization}
\end{minipage}
\vspace{-18pt}
\end{table}
}

\newcommand{\repsunaffiliatedpolarization}{
\begin{table}[htbp]
\centering
\setlength{\tabcolsep}{4pt} % Reduce column spacing
% First tabular (SAGE keywords)
\hspace{-20pt}
\begin{minipage}[t]{0.48\linewidth}
\centering
\tiny
\begin{tabular}{@{}llll@{}}
\toprule
\multicolumn{4}{c}{\textbf{Distinct Keywords, Reps and Unaff-R (Polarization)}} \\ \midrule
{\color{RepublicanRed} \textbf{SAGE}} & {\color{RepublicanRed} \textbf{Republican}} & {\color{UnaffiliatedBrown} \textbf{Unaffiliated}} & {\color{UnaffiliatedBrown} \textbf{SAGE}} \\
0.8192 & \textit{gun} & BPD & 0.8995 \\
0.5704 & business & struggling & 0.7006 \\
0.5016 & white & ADHD & 0.6650 \\
0.5016 & black & diagnosis & 0.5825 \\
0.4774 & jobs & partner & 0.5147 \\
0.4365 & married & psychiatrist & 0.5045 \\
0.4092 & \textit{rape} & wanna & 0.5006 \\
0.4064 & gay & trauma & 0.4855 \\
0.4054 & degree & relate & 0.4607 \\
0.4046 & buy & cry & 0.4523 \\
0.3703 & police & crying & 0.4305 \\
0.3639 & ass & suicidal & 0.4238 \\
0.3626 & perhaps & myself & 0.4206 \\
0.3617 & pay & okay & 0.4194 \\
0.3503 & paying & recently & 0.3912 \\ \bottomrule
\bottomrule
\end{tabular}
\vspace{5pt}
\caption{Open vocabulary based key phrases in posts with polarization language, based on SAGE, Republican and Unaffiliated-R. Words from the polarization language lexicon are italicized. Partisan cultural positions and debates about identity are reflected in words used by Republican users.}
\label{tab:sage_keywords_rep_unaffiliated_polarization}
\end{minipage}
\hspace{7pt}
% Second tabular (Psycholinguistic Differences)
\begin{minipage}[t]{0.49\linewidth}
\centering
\tiny
\begin{tabular}{@{}llll@{}}
\toprule
\multicolumn{4}{c}{\textbf{Psycholinguistic Differences, Reps and Unaff-R (Polarization)}} \\ \midrule
\textbf{Dimension} & {\color{RepublicanRed} \textbf{Republican Mean}} & {\color{UnaffiliatedBrown} \textbf{Unaffiliated Mean}} & \textbf{Adj. p value} \\
{\color{RepublicanRed} \textbf{Social}} & 15.44\% $\pm$ 11.28\% & 14.43\% $\pm$ 10.45\% & $3.82 \times 10^{-170}$ \\
{\color{RepublicanRed} \textbf{Male}} & 1.34\% $\pm$ 3.54\% & 1.19\% $\pm$ 3.21\% & $8.52 \times 10^{-44}$ \\
{\color{RepublicanRed} \textbf{Female}} & 1.57\% $\pm$ 4.22\% & 1.32\% $\pm$ 3.68\% & $1.86 \times 10^{-83}$ \\
{\color{UnaffiliatedBrown} \textbf{Cognitive Processing}} & 15.61\% $\pm$ 9.99\% & 16.34\% $\pm$ 9.63\% & $1.84 \times 10^{-108}$ \\
{\color{UnaffiliatedBrown} \textbf{Positive Emotion}} & 5.11\% $\pm$ 5.88\% & 5.36\% $\pm$ 5.73\% & $1.09 \times 10^{-39}$ \\
{\color{RepublicanRed} \textbf{Negative Emotion}} & 7.77\% $\pm$ 8.98\% & 7.45\% $\pm$ 8.30\% & $9.92 \times 10^{-29}$ \\
{\color{RepublicanRed} \textbf{Anxiety}} & 1.02\% $\pm$ 2.71\% & 1.15\% $\pm$ 2.59\% & $1.17 \times 10^{-43}$ \\
{\color{RepublicanRed} \textbf{Anger}} & 3.46\% $\pm$ 7.05\% & 2.88\% $\pm$ 6.23\% & $3.04 \times 10^{-150}$ \\
{\color{UnaffiliatedBrown} \textbf{Feel}} & 1.34\% $\pm$ 2.90\% & 1.66\% $\pm$ 3.01\% & $2.90 \times 10^{-232}$ \\
{\color{RepublicanRed} \textbf{Bio}} & 5.04\% $\pm$ 7.15\% & 4.81\% $\pm$ 6.57\% & $2.26 \times 10^{-23}$ \\
{\color{RepublicanRed} \textbf{Body}} & 1.47\% $\pm$ 3.83\% & 1.28\% $\pm$ 3.36\% & $5.96 \times 10^{-61}$ \\
{\color{UnaffiliatedBrown} \textbf{Health}} & 1.85\% $\pm$ 3.56\% & 2.03\% $\pm$ 3.58\% & $1.45 \times 10^{-52}$ \\
{\color{UnaffiliatedBrown} \textbf{Affiliation}} & 2.29\% $\pm$ 4.02\% & 2.34\% $\pm$ 3.85\% & $2.15\times 10^{-4}$ \\
{\color{RepublicanRed} \textbf{Work}} & 2.97\% $\pm$ 4.90\% & 2.56\% $\pm$ 4.24\% & $7.38 \times 10^{-161}$ \\
{\color{RepublicanRed} \textbf{Leisure}} & 1.18\% $\pm$ 2.98\% & 1.10\% $\pm$ 2.76\% & $6.44 \times 10^{-17}$ \\
{\color{RepublicanRed} \textbf{Home}} & 0.45\% $\pm$ 1.71\% & 0.41\% $\pm$ 1.49\% & $2.82 \times 10^{-12}$ \\
{\color{RepublicanRed} \textbf{Money}} & 0.89\% $\pm$ 2.73\% & 0.64\% $\pm$ 2.14\% & $3.53 \times 10^{-217}$ \\
{\color{RepublicanRed} \textbf{Religion}} & 0.35\% $\pm$ 1.90\% & 0.26\% $\pm$ 1.52\% & $1.40 \times 10^{-58}$ \\
{\color{RepublicanRed} \textbf{Death}} & 0.43\% $\pm$ 2.09\% & 0.42\% $\pm$ 1.91\% & 0.013 \\
\bottomrule
\end{tabular}
\vspace{5pt} 
\caption{Significant LIWC Dimensions, Polarization Differences, {\color{RepublicanRed}Republican} and {\color{UnaffiliatedBrown}Unaffiliated-R} users. Posts from Republican users use significantly less affiliation language than unaffiliated users.}
\label{tab:liwc_table_rep_unaffiliated_polarization}
\end{minipage}
\vspace{-18pt}
\end{table}
}

\newcommand{\demsunaffiliatedpolarization}{
\begin{table}[htbp]
\centering
\setlength{\tabcolsep}{4pt} % Reduce column spacing
% First tabular (SAGE keywords)
\hspace{-20pt}
\begin{minipage}[t]{0.48\linewidth}
\centering
\tiny
\begin{tabular}{@{}llll@{}}
\toprule
\multicolumn{4}{c}{\textbf{Distinct Keywords, Dems and Unaff-D (Polarization)}} \\ \midrule
{\color{DemocratBlue} \textbf{SAGE}} & {\color{DemocratBlue} \textbf{Democrat}} & {\color{UnaffiliatedBrown} \textbf{Unaffiliated}} & {\color{UnaffiliatedBrown} \textbf{SAGE}} \\
1.475 & trump & BPD & 0.9829 \\
0.4874 & certainly & wanna & 0.6161 \\
0.4808 & \textit{bullshit} & cry & 0.5959 \\
0.4764 & white & struggling & 0.5690 \\
0.4545 & power & relate & 0.5568 \\
0.4382 & business & trauma & 0.5481 \\
0.4356 & black & anxious & 0.5125 \\
0.4309 & society & suicidal & 0.4753 \\
0.4154 & insurance & myself & 0.4730 \\
0.4026 & plenty & tired & 0.4496 \\
0.3973 & line & mood & 0.4403 \\
0.3823 & number & recently & 0.4402 \\
0.3696 & behavior & panic & 0.4346 \\
0.3665 & gay & felt & 0.4186 \\
0.3665 & police & feel & 0.4057 \\ \bottomrule
\bottomrule
\end{tabular}
\vspace{5pt}
\caption{Open vocabulary based key phrases from polarization posts, based on SAGE, {\color{DemocratBlue}Democrat} and {\color{UnaffiliatedBrown}Unaffiliated-D} users. Words from the polarization lexicon are italicized. We observe aspects of structural and economic approaches to mental health reflected in language used distinctly from Democrat users.}
\label{tab:sage_keywords_dem_unaffiliated_polarization}
\end{minipage}
\hspace{7pt}
% Second tabular (Psycholinguistic Differences)
\begin{minipage}[t]{0.49\linewidth}
\centering
\tiny
\begin{tabular}{@{}llll@{}}
\toprule
\multicolumn{4}{c}{\textbf{Psycholinguistic Differences, Dems and Unaff-D (Polarization)}} \\ \midrule
\textbf{Dimension} & {\color{DemocratBlue} \textbf{Democrat Mean}} & {\color{UnaffiliatedBrown} \textbf{Unaffiliated Mean}} & \textbf{Adj. p value} \\
{\color{DemocratBlue} \textbf{Social}} & 14.92\% $\pm$ 11.00\% & 14.16\% $\pm$ 10.46\% & $7.62 \times 10^{-92}$ \\
{\color{DemocratBlue} \textbf{Male}} & 1.22\% $\pm$ 3.40\% & 1.18\% $\pm$ 3.21\% & $5.40 \times 10^{-4}$ \\
Female & 1.36\% $\pm$ 3.93\% & 1.36\% $\pm$ 3.78\% & 0.881 \\
{\color{UnaffiliatedBrown} \textbf{Cognitive Processing}} & 15.86\% $\pm$ 10.03\% & 16.40\% $\pm$ 9.59\% & $6.84 \times 10^{-57}$ \\
{\color{UnaffiliatedBrown} \textbf{Positive Emotion}} & 5.15\% $\pm$ 5.81\% & 5.30\% $\pm$ 5.70\% & $3.09 \times 10^{-14}$ \\
{\color{DemocratBlue} \textbf{Negative Emotion}} & 7.74\% $\pm$ 9.00\% & 7.40\% $\pm$ 8.12\% & $1.39 \times 10^{-29}$ \\
{\color{UnaffiliatedBrown} \textbf{Anxiety}} & 0.99\% $\pm$ 2.63\% & 1.15\% $\pm$ 2.56\% & $1.41 \times 10^{-67}$ \\
{\color{DemocratBlue} \textbf{Anger}} & 3.52\% $\pm$ 7.14\% & 2.94\% $\pm$ 6.27\% & $1.63 \times 10^{-141}$ \\
{\color{UnaffiliatedBrown} \textbf{Feel}} & 1.32\% $\pm$ 2.85\% & 1.68\% $\pm$ 3.06\% & $3.60 \times 10^{-261}$ \\
{\color{DemocratBlue} \textbf{Bio}} & 5.19\% $\pm$ 7.29\% & 4.84\% $\pm$ 6.61\% & $4.21 \times 10^{-48}$ \\
{\color{DemocratBlue} \textbf{Body}} & 1.51\% $\pm$ 3.91\% & 1.36\% $\pm$ 3.61\% & $4.00 \times 10^{-33}$ \\
{\color{UnaffiliatedBrown} \textbf{Health}} & 1.85\% $\pm$ 3.56\% & 1.95\% $\pm$ 3.44\% & $8.09 \times 10^{-15}$ \\
{\color{UnaffiliatedBrown} \textbf{Affiliation}} & 2.21\% $\pm$ 3.90\% & 2.29\% $\pm$ 3.81\% & $1.86 \times 10^{-10}$ \\
{\color{DemocratBlue} \textbf{Work}} & 2.98\% $\pm$ 4.85\% & 2.51\% $\pm$ 4.20\% & $2.74 \times 10^{-195}$ \\
{\color{DemocratBlue} \textbf{Leisure}} & 1.13\% $\pm$ 2.89\% & 1.10\% $\pm$ 2.72\% & $2.38 \times 10^{-3}$ \\
Home & 0.43\% $\pm$ 1.65\% & 0.42\% $\pm$ 1.52\% & 0.179 \\
{\color{DemocratBlue} \textbf{Money}} & 0.85\% $\pm$ 2.70\% & 0.63\% $\pm$ 2.12\% & $4.80 \times 10^{-174}$ \\
{\color{DemocratBlue} \textbf{Religion}} & 0.32\% $\pm$ 1.84\% & 0.26\% $\pm$ 1.55\% & $1.45 \times 10^{-25}$ \\
Death & 0.39\% $\pm$ 1.94\% & 0.40\% $\pm$ 1.86\% & 0.175 \\
\bottomrule
\end{tabular}
\vspace{5pt} 
\caption{Significant LIWC Dimensions, Polarization Differences, {\color{DemocratBlue}Democrat} and {\color{UnaffiliatedBrown}Unaffiliated-D} users. Similar to Republican users, posts from Democrat users use less affiliation language than unaffiliated users.}
\label{tab:liwc_table_dem_unaffiliated_polarization}
\end{minipage}
\vspace{-18pt}
\end{table}
}

\section{Introduction}
\begin{quote}
    ``My [psychotherapy] patients, regardless of political affiliation, are incorporating the messages of social movements into the very structure of their being. New words make new thoughts and feelings possible. As a collective we appear to be coming around to the idea that bigger social forces run through us, animating us and pitting us against one another, whatever our conscious intentions. To invert a truism, the political is personal.''---\textit{Orna Guralnik}~\cite{guralnik2023couples} 
\end{quote}

Mental health concerns are on the rise globally---1 in every 2 people will experience a mental health disorder over the course of their lifetime~\cite{mcgrath2023age}. Intersections of culture, identity, and environment will play a key role in how each individual comes to understand their distress and finds care~\cite{goldberg1980mental, pendse2021can, bhui2002mental}. Technology can play a major role in this process, including enabling individuals to seek information about their distress~\cite{rochford2023leveraging}, build solidarity with others who have similar experiences~\cite{schaadhardt2023laughing}, and find support they may not otherwise have access to due to stigma and marginalization~\cite{feuston2022you, pendse2021can}. Work in Computer Supported Cooperative Work (CSCW)~\cite{pendse2019cross, de2017gender, li2016sunforum, chopra2021living} and Human-Computer Interaction (HCI)~\cite{schaadhardt2023laughing, pruksachatkun2019moments, feuston2022you, milton2023see} has leveraged analyses of online interactions to shed light on the sociocultural aspects of mental health experience. Work in CSCW and HCI has also investigated the core role that online interactions can play in building solidarity among diverse and often marginalized people~\cite{de2016social, wulf2013fighting, crivellaro2014pool, dosono2020decolonizing, simpson2023hey, schaadhardt2023laughing, li2018slacktivists}.

The process of building solidarity across groups can be extremely complex, particularly given rising social division~\cite{devlin2021people} and political polarization~\cite{mccoy2022reducing} globally. In the United States (U.S.), one consequence of extreme political polarization between the two dominant political parties~\cite{dunn2020few, iyengar2015fear, ruckelshaus2022kind, mason2018losing, shafranek2020political, brown2021measurement} has been the creation of unique partisan subcultures associated with each party~\cite{layman1999culture, hetherington2018prius}. Consistent cultural differences between American Republicans and Democrats have been observed to impact non-political spaces, including the cars people buy, media consumed, and preferred choice of caffeinated beverage~\cite{hetherington2018prius, ding2023same, tripodi2022propagandists}, as well as more consequential decisions, such as openness to seeking care for illness~\cite{wallace2023excess, engel2022partisan} and policy positions on the rights of minority groups~\cite{layman1999culture}. Analyses of language utilized by Republicans and Democrats on social media have demonstrated that these cultural differences transfer to online contexts~\cite{ding2023same, sylwester2015twitter, darian2023competing}, including strong ties between partisan culture and how politicians represent themselves~\cite{shi2017cultural}. 

Researchers have argued that rigid partisan cultures have played a role in why American polarization proves difficult to mitigate~\cite{iyengar2019origins, hetherington2018prius}, and that partisan culture plays a core role in how voters perceive policy and politicians~\cite{hiaeshutter2023cued, layman1999culture}. Partisan culture thus has direct ties to policy and power in the U.S. The growing role of partisan culture in how individuals navigate their worlds thus raises potent questions about the potential influence of partisan culture on mental health experience, particularly given the rising prevalence of mental health disorders~\cite{mcgrath2023age}. Mental health policy reform has historically had some bipartisan support from politicians~\cite{goss2015defying}, but in a time of increased cultural polarization, it is unclear if partisan culture has had an influence on how Republicans and Democrats with lived experience of mental illness \textit{express} distress. The type of language that individuals use around their distress can have a strong impact on the type of healthcare they receive~\cite{nichter2010idioms, bhui2002mental, kleinman2020illness} and the type of diverse social connections they are able to form~\cite{gobodo2004human}. Understanding where partisan culture may influence expressions of distress could shed light on whether mental health experiences are another dimension of increasing polarization in the U.S., and whether there may be differences in the type of healthcare people of different partisan identities can access.

Past work in CSCW has examined the role of different aspects of culture in online mental health expressions, including cultural differences from users in countries diverse as India, the United Kingdom, South Africa, Malaysia, the Philippines, and China~\cite{pendse2019cross, de2017gender, zhang2018online}. In this study, we turn this lens to partisan culture in the United States, examining where partisan culture may have an impact on how Republicans and Democrats express distress online, with potential implications for clinical care-seeking and solidarity building. We examine expressions of distress among Republicans and Democrats due to the power they hold in the two-party American electoral system and policy reform~\cite{woolley2023american, engs2002birth, kazin2022took}. We ask the following research question: \textbf{are there partisan cultural differences in how Republican and Democrat users of mental health support communities express distress online?} To answer this question, we utilize a statistical matching method, originating in the causal inference literature~\cite{rosenbaum1985constructing, stuart2010matching} to create a statistically matched sample of 8,916 Republican, Democrat, and unaffiliated Reddit support community members. We then utilize methods from Natural Language Processing (NLP) to analyze a total of 2,184,356 posts from these users spanning spanning January 2013 to December 2022, investigating linguistic differences in mental health expressions. In particular, we examine how partisan cultural norms may be embedded in expressions of distress, including expressions of distress that specifically use mental health language or language indicative of political polarization. 

We find differences in how Republican and Democrat users express distress, including significantly higher language around clinical illness and polarization from Democrat users, and significantly higher language indicative of social relationships and life experiences among Republican users. We build on our findings to discuss implications for how technology designers can support culturally sensitive care and solidarity building among Republicans and Democrats online. 

\section{Related Work}
Below, we provide context on the growth of partisan culture in the U.S. and its implications for policy. We then discuss methods from computing research that utilize language as a lens into mental health experience, cultural difference, and social change.

\subsection{Partisan Culture and Mental Health in the United States}
In this work, we understand culture to be an abstraction used to describe intersections of shared values, norms, worldviews, ideologies, and practices among a group of people. We build on Kleinman and Benson's~\cite{kleinman2006anthropology} clinical definition of culture as a process that is ``inseparable from economic, political, religious, psychological, and biological conditions.''

In the U.S., affiliation with a political party was long associated with belief or advocacy for a set of policy positions~\cite{ruckelshaus2022kind}. However, in the 1970s, affiliation with the Republican or Democratic Party quickly became a \textit{cultural} affiliation alongside a political one~\cite{layman1999culture, hetherington2018prius}, buoyed by an increased religious and racial homogenization of the Republican Party in the decade prior~\cite{maxwell2019long}. For many, partisan affiliation became a representation of broader worldview, belief systems, and values, rather than the ``[bundler] of policy positions''~\cite{ruckelshaus2022kind} it had once been. Republican partisan cultural values developed into being strongly tied to what Hetherington and Weiler~\cite{hetherington2018prius} call \textit{fixed} cultural worldviews, or being ``warier of social and cultural change and hence more set in their ways, more suspicious of outsiders, and more comfortable with the familiar and predictable.'' Democratic partisan culture developed into centering what Heatherington and Weiler dub \textit{fluid} cultural worldviews, or support of ``changing social and cultural norms, [excitement] by things that are new and novel, and [openness]'' around diversity. The impact of these cultural difference have grown over time---Republicans and Democrats diverge in the cars they drive, their choice of caffeinated beverages, the media they consume, the names they give their children, the amount of money they give to charity, levels of religiosity, affiliation toward military culture, and gun ownership~\cite{hetherington2018prius, oliver2016liberellas, margolis2017partisan, betros2001political, parker2017america, schoenmueller2022polarized}. Partisan affiliation has an impact on decisions such as where individuals live, who they choose to marry, and who they vote for, with these impacts being observed even when accounting for interactions between race and gender identity, and independent of policy positions or ideology~\cite{brown2021measurement, green2004partisan}. 

Partisanship has shifted from being organized around ideology to being organized around \textit{culture}, reflecting not only ideology, but also shared worldviews and practices among a group of people. This can be observed in the evolving ideologies (but static cultural identities) among American political parties over the last decade, such as the Republican Party’s move from advocating free trade and strategic international alliances to embracing protectionist trade policies and skepticism of international institutions, while still maintaining consistent cultural identities around rural values, traditional social structures, and patriotic symbolism~\cite{keser2024partisanship, hetherington2018prius}. Similarly, research from the Pew Research Center has described a realignment of the demographics of both Republican and Democrat partisans over the last decade, such as a shift towards Republican partisan identity by racial minority groups~\cite{pew2024changing}. This untethering of partisan affiliation from ideology, demographics, or policy preferences demonstrates how affiliation with the Republican Party or Democratic Party has become a distinct cultural identity, and why we choose to study partisan culture rather than ideological affiliations. 

Partisan cultural differences between Republicans and Democrats have been observed to translate to online contexts that are not explicitly political. For example, Shi et al.~\cite{shi2017cultural} find strong patterns of shared Twitter followers between partisan politicians and popular cultural figures or institutions that are aligned in partisan identity. This includes lifestyle areas as diverse as music, Fortune 100 companies, and restaurant preferences. Similarly, Darian et al.~\cite{darian2023competing} find partisan differences in the kinds of language that conservative Republican and liberal Democrat American bloggers (determined by their sponsoring advocacy organization) use around data. Tripodi~\cite{tripodi2022propagandists} describes partisan cultural differences in how Republicans approach and use Internet resources, such as leveraging methods of analysis taught in Bible study classes (or what Tripodi dubs \textit{scriptural inference}) to analyze the truthfulness of Google Search results. Partisan culture has an impact on online expression broadly, but it is an open question as to how partisan culture might interact with mental health online. 

Understanding partisan differences in mental health language has implications for connecting people to culturally sensitive care, as well as opportunities for social movement building across partisan groups. For example, past research has linked cultural differences in how people conceptualize and express their mental health online to the types of mental healthcare they end up pursuing, in both online~\cite{pendse2023marginalization, kruzan2022wanted} and offline~\cite{kleinman2020illness, nichter2010idioms, bhui2002mental} environments. Work from conflict and reconciliation studies (such as that of Gobodo-Madikizela~\cite{gobodo2004human}) and from political science~\cite{kalla2020reducing, broockman2016durably, kalla2023narrative} has similarly described how connecting across diverse experiences of distress could provide coalition-based foundations for social change. In line with Kleinman and Benson's~\cite{kleinman2006anthropology} call to be attentive to how a patient ``understands, feels, perceives, and responds to [their illness],'' we utilize computational methods to understand how partisan culture influences people's expressions of distress online. 

\subsection{Culture, Social Change, and Mental Health Language Online}
Language can be key to understanding both individual experience of mental health, as well as the social and political alignments of an individual. In the field of medical anthropology, language around mental health that is shaped by intersections between identity, culture, and environment has been dubbed the \textit{idiom of distress}~\cite{nichter2010idioms}. Similarly, in the field of sociology, the social or political alignment associated with terms that an individual chooses to use in non-political spaces has been dubbed the \textit{ideological dialect}~\cite{tripodi2022propagandists}. Tripodi cites the example of the phrases ``undocumented immigrants'' and ``illegal aliens''---two phrases that describe the same group of people but indicate vastly different ideological alignments and worldviews. Language has been demonstrated to play a core role in how people access treatment for mental illness~\cite{snowden1999african, sentell2007access, norris2008avoidance, bhui2002mental}. 

Work in CSCW and HCI has utilized computational methods to analyze cultural diversity in idioms of distress online. For example, De Choudhury et al.~\cite{de2017gender} utilize the Linguistic Inquiry and Word Count (LIWC) text analysis tool to find patterns in differences between how Twitter users in India, South Africa, the U.S., and the UK describe their mental health experiences, tying observed differences to anthropological research on cultural differences. Similarly, Pendse et al.~\cite{pendse2019cross} utilize unigram counts of clinical mental health language derived from the most used terms in the Diagnostic and Statistical Manual of Mental Disorders (DSM)~\cite{american2013diagnostic} to evaluate differences in how Indian, Malaysian, and Filipino Talklife users discuss their mental health on the forum. Users from countries in the Global South were found to use lower levels of negative emotion in their expressions of distress~\cite{de2017gender}, as well as lower levels of clinical language~\cite{pendse2019cross}. Looking to U.S. contexts, Pendse et al.~\cite{pendse2023marginalization} utilize LIWC in combination with a Sparse Additive Generative (SAGE)~\cite{eisenstein2011sparse} language modeling approach to understand differences in how Twitter users in U.S. Mental Health Professional Shortage Areas distinctly describe their distress.       

Similar methods of linguistic analysis have been used as a lens to examine ideological alignment within social movements online. Rho et al.~\cite{rho2018fostering} utilize an LIWC-based approach to analyze comments on Facebook posts from media outlets discussing the \#MeToo movement. Similarly, De Choudhury et al.~\cite{de2016social} utilize an LIWC-based analysis of posts during specific protests within the Black Lives Matter movement, exploring how people's online language changes over time as they engage with the movement. Looking to debates around abortion rights on Twitter, Sharma et al.~\cite{sharma2017analyzing} utilize a SAGE language modeling~\cite{eisenstein2011sparse} approach to find that the ideological stance of individuals can be identified by the unique hashtags they utilize when discussing abortion. Across diverse social movements, online language can be a useful insight into potential places for solidarity building. 

The LIWC text analysis tool~\cite{pennebaker2015development}, often combined with the use of SAGE language modeling~\cite{eisenstein2011sparse}, is well validated as a method to find patterns in differences in online language across studies of partisan difference~\cite{sylwester2015twitter, rho2018fostering}, ideological difference~\cite{sharma2017analyzing, de2016social}, and cultural difference in mental health language~\cite{pendse2023marginalization, pruksachatkun2019moments, de2017gender}. We thus utilize a similar method to analyze how linguistic differences between partisan users of mental health communities may be reflective of underlying cultural differences.

\section{Data and Analytic Approach}
\matchingtable
\matchingfigure
Online mental health communities are largely pseudonymous spaces, to allow for users to comfortably make explicit self-disclosures~\cite{de2014mental}. To identify partisan community members of online mental health communities, we utilized a well-validated method from past studies in computing~\cite{rajadesingan2021political, an2019political, soliman2019characterization}. We identified users who have posted in both political subreddits and mental health subreddits, utilizing validated lists of both types of subreddits from past computing research~\cite{hofmann2022reddit, sharma2018mental}. We then analyzed their posts in mental health communities, understanding these posts to be idioms of distress, in line with past HCI research~\cite{pendse2023marginalization}. Below, we describe our approach in detail.

\subsection{Identifying Partisan and Mental Health Subreddits}
To find partisan users, we utilized a similar method to that of Rajadesingan et al.~\cite{rajadesingan2021political}, An et al.~\cite{an2019political}, and Soliman et al.~\cite{soliman2019characterization}, identifying users based on posting habits in partisan subreddits. We began by utilizing Hofmann et al.'s large scale mapping of partisan subreddits across Reddit~\cite{hofmann2022reddit}, adding to our list any subreddits that the authors validated as politically leaning U.S. Republican or U.S. Democrat. We then added 4 additional subreddits that have been validated in past computing research~\cite{rajadesingan2021political} as being home to either U.S. Democrat or Republican partisan users. Combining these two sets, we arrived at 67 partisan subreddits. 

Past work has described how partisan culture can include electoral norms, or shared belief in electing a specific candidate, which many of the subreddits in Hofmann et al.'s~\cite{hofmann2022reddit} set are centered around. However, partisan culture can also include ideological norms, or shared worldviews and values among people of the same partisan identity~\cite{hetherington2018prius}, alongside other cultural aspects. To find a more diverse set of users under this broader partisan cultural umbrella, we expanded Hofmann et al.'s~\cite{hofmann2022reddit} list, identifying the top 10 subreddits with the highest level of user overlap for those subreddits. We followed Trujillo et al.'s~\cite{trujillo2022make} propensity scoring method~\cite{SubredditStats2023} that measures how likely users from one subreddit are to post or comment in our selected partisan subreddits compared to a randomly selected Reddit user. We took the top 10 subreddits with the highest user overlap score and added them to our list of partisan subreddits, arriving at a list of 193 partisan subreddits after removing duplicates. We conduct a similar process to create our list of mental health subreddits, as past research has discussed how individuals in under-resourced areas may engage with online resources that are not \textit{explicitly} centered on clinical mental health diagnoses~\cite{rochford2023leveraging}. To create this list, we utilized Sharma and De Choudhury's~\cite{sharma2018mental} validated list of 55 mental health subreddits, finding the top 10 subreddits with the highest user overlap for each mental health subreddit. Using this method, we arrived at 199 mental health subreddits after removing duplicates. 

To filter this list, the first and second authors manually reviewed all subreddits, iteratively labeling subreddits that were explicitly related to mental health support or partisan identity, and resolving disagreements until interrater reliability (measured by Cohen's $\kappa$) was high. For political subreddits, this took two rounds, with a Cohen's $\kappa=.79$ on the first round of labeling, and $\kappa=.91$ on the second round after discussion. For mental health subreddits, Cohen's $\kappa$ was $.97$ after the first round of labeling. This process resulted in 69 mental health subreddits (e.g. /r/SuicideWatch or /r/Anxiety) and 94 politics subreddits in total. Reflecting the ideological roots of American partisan culture, it is natural that some of the subreddits that partisan users post in would be related to ideology---we observe overlap in our expanded subreddit list with some ideologically-focused subreddits (such as \textit{/r/socialism}), and our list of subreddits is a reflection of this close tie between partisan culture and ideology. The partisan leaning of newly added subreddits was determined based on whether they or their most overlapped subreddit from past research were coded as U.S. Republican or U.S. Democrat in that past research~\cite{hofmann2022reddit, rajadesingan2021political}. Additionally, through this expansion process, we were able to include mental health subreddits that may not be explicitly centered on mental health diagnoses but include expressions of distress, such as /r/vent or /r/needafriend. A full list of our mental health and political subreddits (including their label as Republican or Democrat, and whether the aspects of partisan culture embodied is primarily ideological or electoral) can be found in Supplements A, B, and C respectively. We highly encourage researchers who are interested in conducting research examining expressions of distress or political partisanship online to leverage these subreddit lists.

\subsection{Finding Partisan Users and Mental Health Posts}
To validate user partisanship and filter out potential trolls, we followed Rajadesignan et al.'s~\cite{rajadesingan2021political} method. We identified Republican and Democrat users that have been active in mental health support communities if all of the following conditions were satisfied: 
\begin{enumerate}
    \item The user has more posts in subreddits associated with a given party than those with the opposing party.
    \item The mean karma score\footnote{On Reddit, a user's ``karma score'' is an ``the upvotes [users] get on [their] posts and comments minus the downvotes''~\cite{encyclopaediaredditica2021}. It is often used in computational social science and HCI research as a measurement of engagement, credibility, or social capital of an individual Reddit user within a Reddit community~\cite{rajadesingan2021political}.} of their posts in subreddits affiliated with the above party is higher than their score in subreddits affiliated with the opposing party. 
    \item The mean karma score in subreddits affiliated with the above party is greater than 1. As Rajadesingan et al.~\cite{rajadesingan2021political} note, 1 is the default karma score than any comment on Reddit receives. A higher than 1 mean score thus indicates net approval from the community. 
    \item The user has posted or commented in a mental health subreddit. 
\end{enumerate}

This method led us to 8,916 Republican users and 26,578 Democrat users active in mental health subreddits. To limit the impact of identity-based dimensions of experience or the imbalance in the number of Republican and Democrat users who actively use mental health forums as confounding factors, following past work from CSCW~\cite{saha2017modeling, ernala2017linguistic, phadke2021makes} that leverages methods from causal inference~\cite{rosenbaum1985constructing, stuart2010matching}, we use statistical matching on user activity and language-based covariates to closely match Republican and Democrat users. We use metrics linked to demographic attributes, personality types, and platform usage as proxy covariates to create a set of Republican, Democrat, and Unaffiliated users that are as similar as possible, to limit the amount of influence that these attributes might have on any partisan linguistic differences we observe. We conduct this matched analysis in light of research demonstrating a realignment of both Republican and Democrat partisans over the last decade, such as shifts towards the Republican Party from racial minority groups~\cite{pew2024changing}. As both American partisan groups are increasingly demographically heterogeneous, our matching process allows for a more comparable sample towards examining partisan cultural differences. We list the covariates we utilized for this analysis below:
\begin{itemize}
    \item \textit{Number of Subreddits}: Users are matched based on the number of subreddits they have posted in, to control for the breadth of their Reddit engagement. 
    \item \textit{Posts Per Month}: Average posts per month are used as a covariate to match users who have similar levels of Reddit engagement. We utilize both the mean and standard deviation of monthly posts as covariates.  
    \item \textit{Length of Post}: The mean and standard deviation length of a user's posts are used as a covariate to match users who have similar levels of content contribution. 
    \item \textit{Function Words}: The Linguistic Inquiry and Word Count~\cite{pennebaker2015development} function word category has been found to work well as a proxy for a range of demographic attributes, identities, and personality types~\cite{chung2011psychological, newman2008gender}. This makes it well suited as a matching variable to reduce the impact of these factors in spaces where demographic information is not readily available. We thus utilize both the mean and standard deviation function word usage per post as matching covariates.
\end{itemize}

We utilized Mahalanobis distance to find the most closely matched Democrat user for each Republican user, leveraging the Hungarian method~\cite{kuhn1955hungarian} to find optimal matches. This resulted in a close match between user groups, as we display in Table~\ref{tab:matching_statistics}, using the Standard Mean Difference (SMD) and Average Variance Ratio (AVR) to measure fit. User groups were also closely matched with regards to the mental health subreddits posted in by users from each user group, as demonstrated by both a Spearman's $\rho$ rank correlation analysis and Kendall's $\tau$ rank correlation analysis ($p < 10^{-22}$, $\rho = .96$,  $\tau = .84$) of the frequency of posts in each mental health subreddit by the users in our dataset. We then repeat this process with a set of what we dub \textit{unaffiliated} users, or users who are not validated partisans following Rajadesignan et al.'s~\cite{rajadesingan2021political} method, and have not actively posted in political subreddits but have posted in mental health subreddits. These users are unaffiliated with regards to their Reddit account use---using it without strongly affiliating themselves with any partisan subreddits. We gather two separate 1\% random samples of all users who have posted in mental health subreddits and are not in our Republican or Democrat user group. We then conduct an identical matching process to that of our Republican and Democrat users, finding the optimal 1-1 matches for each set of 8916 matched Republican and Democrat users. We then leverage these user groups (Unaffiliated-R and Unaffiliated-D, per the group they are matched to) to demonstrate how Republican and Democrat users differ in idioms of distress from typical mental health community users. We collect and analyze all posts from these four user sets for three comparisons. We use \unaffiliatedword{\textbf{brown}} text color and italicized text for findings that apply to unaffiliated users, \republicanword{\textbf{red}} text color and italicized text for Republican users, and \democratword{\textbf{blue}} text color and italicized text for Democrat users. We use \similarword{\textbf{purple}} text color and italicized text for findings that apply to both Republican and Democrat users. We also include information in text and in tables about the partisan association of each finding, alongside these text color annotations.

\subsection{Analytic Approach}
\analysisfigure
Our analytic approach consists of three scopes of analysis, each providing us complementary information about how partisan culture may play into expressions of distress online. We display the exact number of posts analyzed in each scope of analysis in Table~\ref{tab:matching_statistics}. To contextualize our findings, we provide quotes from user posts that illustrate specific differences or nuances to expressions of distress that we found, paraphrased for sensitivity, following past work studying social media and mental health~\cite{pendse2023marginalization, de2016discovering, sharma2018mental} and approaches~\cite{markham2012fabrication}. All linguistic comparisons were conducted using Welch's t-test. To account for multiple comparisons, we adjusted p-values using the False Discovery Rate method, with statistical significance set at $p < .05$. 

\subsubsection{Expressive Differences}
Past work in HCI studying both online mental health communities~\cite{pendse2023marginalization, pruksachatkun2019moments, de2017gender} and online political groups~\cite{sylwester2015twitter, rho2018fostering, sharma2017analyzing, de2016social} has demonstrated that psycholinguistic attributes and distinctly used keyword phrases can be useful in understanding where identity and culture influence expression. We use two similar methods to understand how matched Republican, Democrat, and unaffiliated users in our dataset uniquely express distress. Through doing so, we are able to shed light on nuances to how Republican and Democrat users express distress, including both in comparison to each other and to a set of users who are not as actively partisan online.

For our first method, we used a language modeling approach to identify what Tripodi calls ``keyword phrases,''~\cite{tripodi2022propagandists} or single words and phrases that are distinct to a partisan group, and represent their worldview. We leveraged Sparse Additive Generative (SAGE) language models~\cite{eisenstein2011sparse} to identify the most distinct words from Republican, Democrat, and unaffiliated users in their posts. SAGE is an unsupervised language model that identifies distinct language used when comparing two clusters, utilizing a self-tuned regularization parameter to identify distinct terms for each cluster. Following past work~\cite{pendse2023marginalization, zhou2023synthetic}, we report the difference in log-frequencies from the modeled lexical distribution as the SAGE score, analyzing the 15 unigrams with the highest level of difference for Republican/Democrat, Republican/Unaffiliated-R, and Democrat/Unaffiliated-D users.

We utilized the Linguistic Inquiry and Word Count (LIWC)~\cite{pennebaker2015development} text analysis tool to understand where there may be psycholingustic patterns in expressions of distress that are potentially indicative of partisan cultural differences. LIWC categories \sachinedit{are} groups of words that represent certain psychological constructs (such as the sentiment associated with an expression of distress) or content themes (such as language indicative of religious belief or social connections). An analysis of linguistic differences in words associated with certain LIWC categories allows us to understand patterns in how Republican, Democrat, and Unaffiliated users express distress. To conduct this analysis, we identified the percentage of a given post from a Republican, Democrat, or Unaffiliated user that utilizes words from each LIWC category. We then made comparisons between the vector of LIWC values for specific categories that could demonstrate a meaningful difference between Republican and Democrat users that is grounded in past literature, conducting this analysis between Republican/Democrat, Republican/Unaffiliated-R, and Democrat/Unaffiliated-D users. We excluded any categories that are considered function word categories, given our usage of the category for user matching. A full listing of the specific LIWC categories that form the basis of our analysis and findings can be found in Supplement D, including how past literature motivated our rationale for selecting each LIWC category for analysis. Hereinafter, we use the phrase ``LIWC dimensions'' to refer to our analysis of the incidence of these specific categories (as listed in Supplement D) in posts from Republican, Democrat, and Unaffiliated users, as each category forms the basis for a dimension of linguistic analysis.

\subsubsection{Clinical Language Differences}
Past work studying expressions of distress in online environments has noted the role that identity and cultural factors can play in whether people utilize clinical mental health language to express their distress~\cite{pendse2019cross, rochford2023leveraging}. For example, Pendse et al.~\cite{pendse2019cross} find lower levels of language from the Diagnostic and Statistical Manual of Mental Disorders (DSM)~\cite{american2013diagnostic} among minoritized users from the Global South in an international online mental health community. This past work has emphasized how one major cultural difference in mental health experiences can be whether people understand their distress to be clinical or psychiatric, with ties to the type of healthcare that people receive~\cite{bhui2002mental}. We thus conduct an analysis to understand whether there are difference in the level and types of clinical language (for example, the types of disorders more distinctly mentioned) used by Republican and Democrat users. 

We operationalized clinical language utilizing two lexicons. First, we created a lexicon of clinical mental health language using similar methods to past work in HCI by Gaur et al.~\cite{gaur2018let} and Pendse et al.~\cite{pendse2019cross}. We gathered the top 300 words from the Diagnostic and Statistical Manual of Mental Disorders (DSM)~\cite{american2013diagnostic}. However, this dataset contains many words that are not explicitly clinical, such as ``people'' or ``life.'' To filter out non-clinical stopwords, we followed a Large Language Model (LLM) based filtering approach. Following past research~\cite{zhao2023new, chew2023llm, ziems2023can, pendse2024quantifying}, we prompt OpenAI GPT-4's~\cite{openai2023gpt4} API to provide all terms related to clinical mental health from our set of words. For consistency, we conduct this process 10 times, calculating the interrater reliability across all iterations using Light's $\kappa$~\cite{hallgren2012computing, light1971measures}, finding a kappa of $0.87$. We then calculated the interrater reliability of the filtered terms with a manual labeling of a 10\% random sample of terms from the first author of the study team, finding a $\kappa$ of $0.78$. We provide the specific prompt we used to filter terms and final list of 82 filtered terms in Supplement E. Example words in this filtered lexicon included ``depression'' and ``trauma.'' 

We then analyzed the incidence of posts that contain clinical language across all 8,916 matched Republican/Democrat, Republican/Unaffiliated-R, and Democrat/Unaffiliated-D users, to see if certain partisan groups have higher levels of clinical language. We then identified matched Republican/Democrat, Republican/Unaffiliated-R, and Democrat/Unaffiliated-D users who have both used clinical mental health language in at least one post to do a focused comparison of posts with clinical language, using a similar psycholinguistic and open vocabulary keyword phrase approach. Specific statistics on the closeness of matches for each category are in Table~\ref{tab:matching_statistics}. 

\subsubsection{Polarization Language Differences}
Past work studying partisan differences online has discussed the role of partisan polarization in online language used by Republicans and Democrats~\cite{rho2018fostering, sylwester2015twitter, tripodi2022propagandists}. We investigate whether Republican and Democrat users distinctly utilize language indicative of polarization (subsequently called polarization language) in their expressions of distress. Through this analysis of polarization language, we shed light on whether political language is embedded in distress language online, and investigate whether further empirical support exists for Kleinman and Benson's~\cite{kleinman2006anthropology} argument that cultural language around distress cannot be separated from political conditions.  

We operationalize polarization language through use of the online polarization language dictionary created by Simchon et al.~\cite{simchon2022troll}, empirically validated across timepoints and social media platforms, including Reddit. Example words in this dictionary include ``crime,'' ``Republicans,'' and ``Democrats,'' with 205 words in the lexicon. We conduct a similar analysis to our clinical analysis, finding matched Republican/Democrat, Republican/Unaffiliated-R, and Democrat/Unaffiliated-D users who both utilize polarization language in at least one post, with matching statistics in Table~\ref{tab:matching_statistics}. We then replicated our clinical analysis for polarization language, testing for differences in incidence of polarization language, distinct keyword phrases, and psycholinguistic patterns. 

\section{Partisan Expressive Differences}
We begin with our analysis of partisan expressive differences, analyzing how Republican, Democrat, and Unaffiliated users distinctly describe their experiences of distress online.

\subsection{Open Vocabulary Expressions}
\expressivedifferencessage
We start by analyzing the keywords, in an open vocabulary format, most distinctly used by Republican and Democrat users, as obtained through use of the SAGE algorithm (ref. Section 3.3.1). We present these findings in Table~\ref{tab:expressive_keyphrases}.

We find that both Republican and Democrat users distinctly use language indicative of occupation, income, and class in their expressions of distress, both using the terms \similarword{business} and \similarword{paying}. In line with this broader theme, Republican users distinctly use the word \republicanword{jobs}, \republicanword{pay}, and \republicanword{buy}, and \republicanword{degree}. We also find expressions of racial, sexual, and gender identity to play a role in partisan expressions of distress, with \similarword{white}, \similarword{black}, and \similarword{gay} all being used distinctly by partisan users, and \democratword{women} being used distinctly by Democrat users. This suggests that both socioeconomic concerns as well as discussions of identity (and potentially associated partisan debates around identity~\cite{lindaman2002issue, AmericanPoliticalScienceAssociation2023}) are a part of expressions of distress for partisans. 

We observe certain themes to be distinctly talked about by Republicans and Democrat users, tied to past research on partisan cultural differences. Democrat users distinctly use language in their expressions of distress indicative of broader social issues, including use of the words \democratword{society}, \democratword{power}, and \democratword{history}. This is in line with Hetherington and Weiler's finding that Democrats tend to be more open to discuss social change, whereas Republicans feel more comfortable with discussions that are local and immediate~\cite{hetherington2018prius}, including in online contexts~\cite{hemsley2021staying}. Compared to their Republican counterparts, Democrat users also distinctly use language that directly relates to clinical mental health, including the terms \democratword{ADHD}, \democratword{diagnosis}, \democratword{dose}, \democratword{brain}, \democratword{psychiatrist}, \democratword{autistic}, and \democratword{diagnosed}. In contrast, Republican users tend to distinctly use language related to day-to-day life experiences, including the word \republicanword{local} and \republicanword{married}. We find this to also be true when compared to Democrat users, with Republicans using \republicanword{local}, 
\republicanword{degree}, \republicanword{dating}, \republicanword{married}, \republicanword{jobs}, and \republicanword{college}. Republican users also distinctly use the word \republicanword{god} when compared to Democrat users and \republicanword{police} when compared to unaffiliated users, which are in line with past work demonstrating strong ties between religion and Republican partisan culture~\cite{schwadel2017republicanization, maxwell2019long, layman1999culture}, and Republican cultural advocacy for law enforcement~\cite{zoorob2019blue}. 

These broader patterns may point to differences in the explanatory models~\cite{kleinman1978culture} that Republicans and Democrats have around their distress---Republicans may see their distress as directly tied to their daily life experiences, whereas Democrats may tie their experiences to psychiatric diagnoses. Alternatively, it may be the case that Democrat users have higher levels of access to language around mental health disorders. This is in line with Rochford et al.'s~\cite{rochford2023leveraging} finding that urban Google search users tend to utilize specific clinical terms like \textit{psychosis} and \textit{ADHD} more actively in web searches than people in rural areas. 

\subsection{Psycholinguistic Differences}
\expressivedifferencesliwc
We observe psycholinguistic differences that hint at different forms of use of online mental health support platforms by Democrat and Republican users. We present these findings in Table~\ref{tab:psycholinguistic_liwc_differences}, with full tables and p-values available in Supplement F. We find consistent differences between partisans and unaffiliated users. We also find that these differences are further accentuated when comparing Republican users and Democrat users---language from Democrat users is largely in line with typical community member expressions of distress, and Republican language in line with partisan trends. 

We find that Republican and Democrat users tend to use significantly more language indicative of their social contexts compared to unaffiliated users. For example, an average post by a Republican user contains 12.93\% ($\sigma = 13.80\%$) social language and an average post from a Democrat user contains 12.32\% ($\sigma = 13.35\%$) social language. In contrast, matched unaffiliated users show significantly lower percentages: 11.85\% ($\sigma = 12.83\%$) and 11.59\% ($\sigma = 12.78\%$) of their average posts, respectively ($p < 10^{-308}$, $p < 10^{-156}$). This pattern extends to language pertaining to gender and gender roles, with Republican and Democrat users utilizing significantly more gender-related language in comparison to unaffiliated matched users ($p < 10^{-115}$ and $p < 10^{-12}$ for male language, and $p < 10^{-259}$ and $p < 10^{-22}$ for female language). Such language often appears in reference to relationships with partners (such as \democratword{``I really miss him...''} or \republicanword{``Should I break up with my \sachinedit{girlfriend?''}}). Further, among partisan users, we observe that Republican users employ more social and gendered language. The average post from a Republican user contains 12.93\% ($\sigma = 13.80\%$) social language, whereas the average post from a Democrat user contains 12.32\% (13.35\%) social language ($p < 10^{-163}$). Illustratively, one Republican user stated: \republicanword{``The depression is difficult, and I've honestly given up on trying to make friends---it gets better? Guys, that's just something normies with lots of friends believe.''} These findings suggest that partisan users, especially those from the Republican side, tend to express their distress by directly speaking about the social relationships in their lives.

Similarly, terminology used to narrate day-to-day life is also employed significantly more frequently by partisan users, with work (R: $p < 10^{-250}$, D: $p < 10^{-308}$), leisure (R: $p < 10^{-12}$, D: $p < 10^{-3}$), home (R: $p < 10^{-33}$, D: $p < 10^{-12}$), money (R: $p < 10^{-308}$, D: $p < 10^{-281}$), and religion (R: $p < 10^{-88}$, D: $p < 10^{-24}$) language significantly higher for Republican and Democrat users compared to their matched unaffiliated users. Compared to Democrat users, we find that Republican users employ language around work ($p < 10^{-3}$), leisure ($p < 10^{-5}$), home ($p < 10^{-15}$), money ($p < 10^{-25}$), and religion ($p < 10^{-29}$) at significantly higher rates. This use of religious language aligns with the strong association between Republican culture and religious belief~\cite{schwadel2017republicanization, maxwell2019long, layman1999culture}, and also indicates that such cultural features influence expressions of distress. For instance, this Republican user comforts another on Easter: \republicanword{``On this blessed day, I'll pray for our well-being, friend.''} 

Politically unaffiliated users tend to use more language that reflects cognitive processing, including 14.69\% ($\sigma = 13.23\%$) of a given post from Unaffiliated-R users, and 14.64\% ($\sigma = 13.20\%$) of a given post from Unaffiliated-D users, compared to 14.26\% ($\sigma = 13.59\%$) from Republican users and 14.57\% ($\sigma = 13.52\%$) from Democrat users. However, this difference in usage is more pronounced when comparing unaffiliated users to Republican users ($p < 10^{-60}$) than Democrat users {$p < .02$}. Among partisans, we find that Democrat users employ more cognitive processing language than Republicans. On average, a Democrat user's post contains 14.57\% ($\sigma = 13.52\%$) cognitive processing language, compared to 14.26\% ($\sigma = 13.59\%$) among Republican users ($p < 10^{-44}$). For example, one Democrat user applies cognitive processing language to describe introspection and uncertainty around their obsessive thoughts: \democratword{\sachinedit{``How} do I stop obsessing and overanalyzing things? It's hard to explain, but I latch on to things and my mind will just continue to repeat itself. I don't understand why this is happening or what would make this better, I just want to be normal.''} The elevated use of cognitive processing language by politically unaffiliated users suggests a key difference in their approach to mental health support communities: they tend to use these platforms to collectively process life events, in contrast to partisan users who more often directly report their life experiences. Further, these results suggest that Democrat users are more inclined to participate in the collective processing of events in these spaces, while Republican users may be seeking a place to vent.

We observe that language around certain emotional states is higher for both groups of partisan users compared to unaffiliated users. We find language indicative of anger and negative emotion to be significantly higher for Democrat and Republican users, and language indicative of positive emotion, broader feelings, and anxiety higher for unaffiliated users. For example, negative emotion language typically comprises 5.36\% ($\sigma = 9.19\%$) and 5.41\% ($\sigma = 9.19\%$) of a given post from a Republican and Democrat user respectively, compared to 5.23\% ($\sigma = 8.67\%$) and 5.15\% ($\sigma = 8.60\%$) from an Unaffiliated-R and Unaffiliated-D user respectively, with $p < 10^{-13}$ and $p < 10^{-45}$. Democrat users, whose linguistic patterns tend to resemble those of unaffiliated users in other respects, diverge from this trend here by showing a usage of emotional language more akin to Republican users. Posts from Democrat users tend to use negative emotion language in 5.41\% ($\sigma = 9.19\%$) of an average post, compared to 5.36\% ($\sigma = 9.19\%$) of a Republican user's average post ($p < 10^{-4}$). Past work utilizing LIWC to analyze social media posts from Democrat and Republican users has observed higher levels of negative emotion, anger, anxiety, and feeling-related language among right-leaning Facebook users discussing the Me Too Movement~\cite{rho2018fostering} and higher levels of feeling and anxiety-related language among the tweets of Democrat Twitter users~\cite{sylwester2015twitter}. In our analysis, we observe higher levels of anger, negative emotion, and feeling-related language among Democrat users, suggesting that there are shifts in partisan cultural norms of expression based on context.  

Compared to unaffiliated users, both groups of partisan users employ higher levels of language related to biological processes and the body in posts. On average, posts from Republican users feature 4.52\% ($\sigma = 8.62\%$) language indicative of biological processes, and posts from Democrats feature 4.60\% ($\sigma = 8.56\%$). Comparatively, matched Unaffiliated-R and Unaffiliated-D users feature 4.39\% ($\sigma = 7.94\%$) and 4.34\% ($\sigma = 8.03\%$) respectively. However, Republicans use significantly less language indicative of health terms than Unaffiliated-R users ($p < 10^{-26}$), employing health language in 1.90\% ($\sigma = 5.38\%$) of a post on average, compared to 2.05\% ($\sigma = 5.20\%$) from unaffiliated users. The two partisan user groups use similar levels of health language. However, Democrats use significantly higher levels of both biological process language ($p < 10^{-8}$), and language related to the body ($p < 10^{-7}$). For example, one Democrat user notes: \democratword{``I'm just too in my head during conversations! I feel like my brain is broken...''} These differences may suggest a greater familiarity with clinical language among Democrat users, in line with our open vocabulary analysis. We delve deeper into this question in our next analysis.

\section{Partisan Differences in Clinical Language}
We now conduct a focused analysis of differences in use of clinical terms across our three user group comparisons, towards understanding whether there are nuances to the level and types of clinical language used across partisan groups, given implications for healthcare utilization and treatment~\cite{bhui2002mental, rochford2023leveraging}.

\subsection{Comparative Utilization of Clinical Language}
We find that unaffiliated users are significantly more likely to use clinical language (and use more clinical language within their posts) compared to partisan users. In the case of our Republican and Unaffiliated-R user comparison, Unaffiliated-R users use clinical language in 20.93\% ($\sigma = 40.29\%$) of posts, compared to 16.43\% ($\sigma = 37.05\%$) of posts from Republican users ($p<10^{-308}$). Looking to language within posts, 1.32\% ($\sigma = 4.38\%$) of an average post from an Unaffiliated-R user contains clinical language, compared to 1.11\% ($\sigma = 4.09\%$) of the average Republican user's post ($p < 10^{-134}$). Similarly, 20.55\% ($\sigma = 40.40\%$) of posts from Unaffiliated-D users contain clinical language, compared to 17.75\% ($\sigma = 38.21\%$) of posts from Democrat users ($p < 10^{-138}$). 1.32\% ($\sigma = 4.37\%$) of a given post from an Unaffiliated-D user uses clinical language, compared to 1.24\% ($\sigma = 4.43\%$) of a given post from a Democrat user ($p < 10^{-19}$). 

Comparing Republican and Democrat users, we find that Democrat users do utilize clinical language significantly more often than Republican users. Democrat users employ clinical language in 17.75\% ($\sigma = 38.21\%$) of posts, whereas Republican users employ clinical language in 16.43\% ($\sigma = 37.05\%$) of posts ($p < 10^{-101}$). Democrat users also use more clinical language within posts in comparison to Republicans, with 1.24\% ($\sigma = 4.43\%$) of a given post from a Democrat user using clinical language, and 1.11\% ($\sigma = 4.09\%$) of a post from a Republican user using clinical language ($p < 10^{-75}$). Our finding recalls previous research which highlights how culturally diverse expressions of distress can go unrecognized by clinicians, and create obstacles in the search for care~\cite{bhui2002mental}. Past work has discussed this phenomenon in the case of racial minorities, but this perspective may also apply to partisan users in their search for care, in that Democrat users may be more able to access care as a result of their use of more clinical language. We investigate the kinds of posts that utilize clinical language among our matched user sets in greater detail below.    
\clinicaldifferencessage
\subsection{Open Vocabulary Expressions}
We now present our analysis of the language that Republican and Democrat users distinctly use alongside clinical terms when expressing distress. We present these findings comparing Republican to Democrat users, Republican users to their unaffiliated counterparts, and Democrat users and their unaffiliated counterparts in Table~\ref{tab:clinical_keyphrases}.

In posts utilizing clinical language, we see similar patterns to our expressive analysis, including a distinct use of clinical terminology from Democrat users. However, terminology in posts with clinical language from Democrat users hints at more social and medical framings of mental illness compared to unaffiliated users, including distinct use of  \democratword{condition(s)}, \democratword{disability}, \democratword{behavior}, \democratword{spectrum}, and \democratword{exercise}. Democrat users also distinctly use the terms \democratword{manage} and \democratword{disability} when compared to Republican users. This may reflect a tendency to view mental health issues through clinical or medical lenses, including how treatment is understood (e.g. exercise). Additionally, the use of the word disability may suggest an explanatory model of illness that frames mental health concerns in terms of chronic conditions that impact an individual's ability to function daily, in line with progressive disability rights and justice frameworks~\cite{burnim2015promise}. 

Terms from Republican users alongside clinical language similarly connect to partisan cultural differences. The most distinctly used word by Republican users when compared to politically unaffiliated online mental health community members is \republicanword{god}, aligning with past work on the religious nature of Republican partisan culture. Work in CSCW has emphasized how important spiritual support can be for individuals in online support communities~\cite{smith2021spiritual}, and this may be particularly true for Republican users of online support communities. We do observe some clinical language being used by Republicans, such as the term \republicanword{psychotic} when compared to unaffiliated users, and \republicanword{schizophrenia}, and \republicanword{PTSD} when compared to Democrat users. In a few cases, these words are used by Republican forum participants as pejoratives (such as describing politicians or people in their life as psychotic). However, Republican users largely use these words to describe their individual experiences of distress or medications they take, tied to the mental health subreddits where our data originates. For example, one Republican user describes how they \republicanword{``managed to take several courses and earn an above-average salary while coping with psychotic episodes,''} encouraging other forum participants to persevere when they take anti-psychotic medications. This suggests an awareness of some diagnostic terms among Republican users, but given the distinct use of \republicanword{hotline} and \republicanword{suicide} by Republican users, this may also suggest a higher level of crisis and greater reliance on non-medical care among Republican users. Counties that are primarily rural tend to both be Republican~\cite{parker2018urban} and have higher suicide rates~\cite{casant2022inequalities, mohatt2021systematic}. Our findings may be a reflection of this reality. 

\subsection{Psycholinguistic Differences}
\clinicaldifferencesliwc
In our psycholinguistic analysis of posts with clinical language, we find similar patterns to our findings in 4.2---Republican users use significantly more language indicative of social relationships and most dimensions of day-to-day life in their posts, whereas Democrat users use language that is indicative of higher levels of cognitive processing. However, there are differences that may shed light on the different lived experiences Democrat users and Republican users have with their mental health. We present our findings for this analysis for Republican and Democrat users, Republican and Unaffiliated-R users, and Democrat and Unaffiliated-D users in Table~\ref{tab:clinical_liwc_differences}, with full tables and p-values available in Supplement F. 

Following the distinct use of \republicanword{suicide} in our analysis of open vocabulary expressions, we observe a greater use of language around death among Republican users when compared to Democrat users. While Republicans discuss death in .47\% ($\sigma = 2.02\%$) of a post on average, Democrat users discuss death in .41\% of a post on average ($\sigma = 1.95\%$), with significance at $p < 10^{-11}$. However, when Republican users are compared to matched unaffiliated users, they use similar levels of language related to death, with a non-significant difference of $p < .08$. Republican discussions of death can be diverse, and often related to suicide. For example, one Republican user emphasizes their social relationships when discussing death: \republicanword{``I promised my dad I will not kill myself until he dies. But it is killing me, all I can think of is suicide.''} Similarly, another Republican user describes their distress within the frame of social comparison: \republicanword{``Why can't I just die? Living people are lucky enough to die all the time.''} These differences in levels of language around death, in combination with the unique usage of the terms \republicanword{suicide} and \republicanword{hotlines} in our open vocabulary analysis, may suggest that Republican users more heavily use online mental health forums for support for suicidal ideation. 

Through our focused analysis of clinical language, we also observe evidence that Republican and Democrat users have differing explanatory models of illness. In our analysis of expressive differences (as presented in Table~\ref{tab:psycholinguistic_liwc_differences}), we observe that usage of language related to the body is significantly higher for Democrat users when compared to Republican users ($p < 10^{-7}$), and usage of language related to health is roughly equal between both Republican and Democrat users ($p < .07$). However, this does not hold when we strictly analyze posts with clinical language. We find that language related to the body is used to a greater extent by Republican users, with 1.43\% ($\sigma = 3.33\%$) of a typical post using body language, compared to 1.36\% ($\sigma = 3.18\%$) from Democrat users ($p < 10^{-6}$). Language related to health is more used in posts by Democrat users, with an average post from a Democrat user utilizing 3.85\% ($\sigma = 5.82\%$) health language, and a typical post from a Republican user containing 3.77\% ($\sigma = 5.87\%$) health language ($p < 10^{-4}$). This difference may be indicative of a split between distress language that is primarily somatic from Republican users, versus distress language that is primarily medical from Democrat users. A Republican user illustrates this point through using physical (body) language to express their distress: \republicanword{``I feel so tired and sick and sleep deprived, I just want to go to sleep forever.''}. Alternatively, a Democrat user uses clinical (health) language related to treatments to describe theirs: \democratword{``Yeah I feel pretty bad now, but I've recognized that medication is only 50\% of the treatment. The other 50\% is learning how to manage my symptoms via therapy.''}. Past work in CSCW has described how people in resource-constrained areas use somatic language to describe their distress on social media~\cite{pendse2023marginalization}---we observe this to similarly be the case for Republican users of online mental health communities. 

\section{Partisan Differences in Polarization Language}
We conclude our analyses by analyzing the incidence and types of political language in expressions of distress by partisan users, to investigate how political factors may become embedded in expressions of distress.
\subsection{Comparative Utilization of Polarization Language}
We present our findings for our analysis of use of language indicative of political polarization across all matched users. We now turn our focus to language indicative of political polarization. We find that both partisan groups are more likely to use polarization language (and higher levels of it within their expressions) when compared to unaffiliated users.

Comparing Unaffiliated-R users and Republican users, Republican users use slightly more than Unaffiliated-R users. Republican users use polarization language in 35.61\% ($\sigma = 47.88\%$) of posts, compared to 35.38\% ($\sigma = 47.82\%$) of posts from Unaffiliated-R users ($p < .02$). Republican users also use more polarization language within their posts, with 3.09\% ($\sigma = 7.28\%$) of a given post from a Republican user comprising polarization language on average, compared to 2.67\% ($\sigma = 6.52\%$) for Unaffiliated-R users ($p < 10^{-192}$). Work studying political polarization on Reddit has discussed how 2016 was a significant moment of polarization~\cite{waller2021quantifying}. We find that prior to 2016, unaffiliated users utilized significantly more polarization language (both in frequency and in percentage of a given post) compared to Republican users, but during and after 2016, Republican users utilized more polarization language. This was not the case for Democrat users, who consistently demonstrated higher levels of polarization language usage both before and after 2016. We list our full results for this temporal analysis in Supplement G. 

We find that Democrat users also use higher levels of polarization language compared to Unaffiliated-D users. Democrat users use polarization language in 37.10\% ($\sigma = 48.31\%$), compared to 35.64\% ($\sigma = 47.89\%$) of posts from Unaffiliated-D users ($p < 10^{-47}$). Democrat users employ more clinical language within their posts, with polarization language taking up 3.22\% ($\sigma = 7.44\%$) of the average post from Democrat users, and 2.69\% ($\sigma = 6.57\%$) of the average post from Unaffiliated-D users ($p < 10^{-278}$).

\sachinedit{When comparing Republican and Democrat users, we find that Democrat users use polarization language more frequently, and to a greater extent, than Republican users.} 37.10\% ($\sigma = 48.31\%$) of posts from Democrat users use polarization language, compared to 35.61\% ($\sigma = 47.88\%$) of posts from Republican users ($p < 10^{-79}$). Similarly, posts from Democrat users tend to use more polarization language on average, with 3.22\% ($\sigma = 7.44\%$) of a given post from Democrats being polarization language on average, and 3.09\% ($\sigma = 7.28\%$) of a given Republican user's post. In line with our open vocabulary SAGE analysis of expressive and clinical differences, this may be the result of a greater integration of political and societal metaphors for illness experience when Democrat users describe their distress. We investigate further by analyzing expressions of distress that have polarization language from our matched user sets below. 

\subsection{Open Vocabulary Expressions}
\polarizationdifferencessage
% \repsunaffiliatedpolarization
% \demsunaffiliatedpolarization
% \repsdemspolarization
We now present our analysis of distinct terms used alongside language that is indicative of political polarization. We present these findings comparing Republican and Democrat users, Republican users to their unaffiliated counterparts, and Democrat users and their unaffiliated counterparts in Table~\ref{tab:polarization_keyphrases}. Similar to our analysis of keywords across all expressions, and our analysis of terms that occur alongside clinical language, we find similar patterns in posts that feature polarization language when comparing Republican and Democrat users, including more psychiatric language (such as \democratword{ADHD} and \democratword{psychiatrist}) among Democrat, and \republicanword{god} among Republican users). However, we do see the term \democratword{trump} uniquely discussed by Democrat users, which is evidence that partisan politics do have an influence on how Democrats discuss their distress. This is in line with Krupenkin et al.'s~\cite{krupenkin2019president} argument that Democrats use political frames to express their distress, particularly to people from similar partisan backgrounds. 

It is when we analyze Republican and Democrat users to unaffiliated users that distinctions between how Republican and Democrat users integrate political language into their expressions of distress become clear. We observe distinct use of the word \republicanword{gun}, the most word used distinctly by Republican users in expressions of distress. Guns are a salient and polarizing symbol in both Republican and Democrat partisan culture~\cite{parker2017america}. Their use in this mental health context may be the influence of guns being a central part of Republican partisan culture~\cite{hetherington2018prius, parker2017america}. For example, Republican forum participants describe specific types of guns that they would like to end their life with, or emphasize the importance of gun ownership towards their personal feelings of safety (e.g. \republicanword{``I started carrying a gun following my assault, and you could too if you're feeling anxious, along with talking to the police''}). The appearance of the word \republicanword{gun} in this context may thus be an impact of the cultural aspects of a high value placed on gun ownership~\cite{boine2020gun} within the daily lives of Republican users with lived experience of mental health needs. Among Democrat users, we observe unique mentions of \democratword{society}, \democratword{power}, and \democratword{trump} similar to our analysis of expressive differences. Given a significantly higher use of polarization language among Democrat users, the appearance of these words may be indicative of Democrat users being more likely to attach political significance to experiences of distress. 

\subsection{Psycholinguistic Differences}
\polarizationdifferencesliwc
In line with our other expressive and clinical psycholinguistic analyses, as presented in Table~\ref{tab:polarization_liwc_differences}, we observe a similar pattern in posts with polarization language---social and day-to-day language is more prevalent among Republican users, and language centered on cognitive processing or biological processes is more prevalent among Democrat users. We turn our attention to the affiliation dimension in LIWC, which includes words that reflect an individual's own description of their connections with communities or group identities~\cite{ashokkumar2022tracking, pennebaker2015development}. Much has been written on how partisanship is a group and social identity~\cite{mason2018losing, mason2018uncivil}, alongside a cultural identity. In posts expressing distress that also feature polarization language, we find that both Democrat users and Republican users use less affiliation language in their posts compared to unaffiliated users. Posts from Republican users utilize 2.29\% affiliation language ($\sigma = 4.02\%$) on average, compared to 2.34\% ($\sigma = 3.85\%$) from matched unaffiliated users ($p < .0002$). Posts from Democrat users similarly use less affiliation language, with Democrat posts featuring 2.21\% ($\sigma = 3.90\%$) affiliation language, and posts from matched unaffiliated users utilizing 2.29\% ($\sigma = 3.81\%$) affiliation language on average ($p < 10^{-10}$). Partisan polarization often manifests as an ``us versus them'' dynamic, but it is possible that in the context of mental health forums, this phenomena is less pronounced, as users do not necessarily report their partisan affiliation in these spaces. When Republican users are compared with Democrat users, Republican users utilize higher levels of affiliation language overall, with 2.29\% ($\sigma = 4.02\%$) from Republican users, and 2.20\% ($\sigma = 3.90\%$) from Democrat users. Hetherington and Weiler~\cite{hetherington2018prius} have discussed how Republican partisan culture is often tied to community contexts and group contexts, which may have an influence on how Republican users are inclined to express distress. As one Republican user put it, \republicanword{``oh, you feel depressed? That makes two of us bud, come join the club.''}.

\section{Discussion}
Partisan polarization has risen steeply in the United States, with fewer and fewer places for people from different partisan backgrounds to connect~\cite{mason2018uncivil}, and a strong influence on expression and rhetoric~\cite{kalmoe2022radical}. We found strong differences in how U.S. Republican and Democrat members of online mental health support communities express their distress, tied to the partisan culture of each group. In this section, we describe the implications of our findings towards culturally sensitive mental health support, and how a consideration of partisan differences in language can support more organized advocacy for policy reform. 

\subsection{Implications for Culturally Sensitive Care}
We found that Democrat users systematically use more clinical language in their mental health expressions online, including distinct mentions of specific psychiatric disorders. Republican users utilize language indicating deeper descriptions of their social interactions and personal experiences. These differences in language may reflect differing cultural uses of online mental health support---Democrat users may approach online mental health communities as a place to learn about and process clinical information, and find community with others who have similar diagnoses. Republican users may utilize online support communities to vent about experiences occurring in their lives that they could not express in offline contexts. Past computing research has discussed the use of online platforms for both information seeking about mental illness~\cite{rochford2023leveraging} as well as for community support~\cite{milton2023see}---our study points to the likelihood that there are partisan cultural differences in how people utilize online support platforms, potentially tied to cultural norms around the understanding and treatment of mental health concerns. Designers of mental health support platforms could build on our finding by creating experiences that accommodate these differing means of use and expression. For example, online mental health spaces for Republican-leaning communities could provide facilitated group discussions or interactive storytelling elements without an explicitly clinical framing. Conversely, it may be more beneficial to a Democrat-leaning community to explore more formal mental health resources and biomedical framings of mental health disorders. Given the potential for identity-focused communities to increase polarization, dialogue would need attentive moderation to ensure that users are respectful and empathetic across partisan boundaries.  

Emerging initiatives could also explore the factors that promote a sense of comfort among Republican users when discussing their mental health, and subsequently establish online spaces that incentivize such discussions. One possible approach involves nudging users to express their distress using language that alludes to models of illness that are more accepted within their specific partisan cultural group. Republican users in our dataset distinctly mentioned PTSD and schizophrenia in our analyses of clinical language, and Democrat users distinctly mentioned ADHD. Additionally, both groups had progressive and conservative political issues as a part of their expressions of distress, such as a focus on power from Democrat users and a focus on God from Republican users. These diagnostic and cultural trends may be an entry point at more targeted support for individuals from differing partisan backgrounds. For example, trauma might be a more accessible way for Republican partisans to discuss their experiences with mental distress, or focusing on social change might effectively encourage a Democrat user to open up about their mental health.     

It is important to emphasize that the patterns we observe may not hold for all Republican and Democrat users, particularly given regional differences in partisan culture~\cite{erikson1987state}. However, our research suggests routes to a greater consideration of a new dimension of culture in mental healthcare, towards (recalling Kleinman and Benson's framing of cultural competency~\cite{kleinman2006anthropology}) being attentive to how a partisan patient ``understands, feels, perceives, and responds to [their illness].''  

\subsection{Implications for Coalition Building and Policy Reform}
In discussing her experiences on South Africa's post-apartheid Truth and Reconciliation Commission, Gobodo-Madikizela~\cite{gobodo2004human} notes the importance of creating spaces where people can express how they have ``lived through pain differently'' utilizing their own language around pain. However, Gobodo-Madikizela also notes the importance of doing so while also supporting ``the forging of a vocabulary of compromise and tolerance'' alongside a consideration of differences, calling this process ``part of the project of creating the operating rules of the democratic game.'' Gobodo-Madikizela argues that the maintenance of a diverse and multicultural democracy in the wake of identity-based violence, conflict, and polarization is dependent on the ``dynamics that impel most of us toward forming an empathic connection with another person in pain, that draw us into his pain, regardless of who that someone is.'' Empirical work in U.S. contexts has demonstrated that shared experiences of distress may help people be more empathetic across cultural and identity-based lines~\cite{kalla2020reducing, broockman2016durably, kalla2023narrative}. The success of this process (dubbed \textit{perspective-getting}~\cite{kalla2023narrative}) is highly dependent on both individuals being able to express their distress in a way that is both true to their own experiences, but also understandable to their conversation partner.

What does the forging of a common vocabulary of compromise and tolerance look like in mental health conversations between historically polarized~\cite{mason2018losing} Republican and Democrat individuals? How can there be compromise and tolerance given continued partisan rhetoric that endorses violence and harm~\cite{kalmoe2022radical}? Our findings point to language (such as that around guns, the police, and religion) that is particularly polarizing in contemporary American society also being used to express distress. However, our findings also point to online mental health communities being one space where diverse people can express how they have ``lived through pain differently,'' given our finding that both Republican and Democrat users utilize online mental health communities for support, though their language and non-anonymous level of use might be different. Our work points to the potential for well-moderated online mental health support spaces to provide rare opportunities for people from different political backgrounds to empathize and connect. This is an especially salient point, for two reasons. The first being, as Robert Putnam noted two decades ago, Americans are increasingly ``bowling alone''~\cite{putnam2000bowling}; thus opportunities where they can physically meet and connect as a community, in spite of their ideological differences, are rapidly shrinking. The second being that, at the same time, recent times have witnessed online communities emerging as virtual ``third places''~\cite{soukup2006computer} -- in sociologist Ray Oldenburg's terms, the public spaces we often use for informal social interaction outside of the home and workplace~\cite{oldenburg1982third}. Careful design could help support online spaces as places for reconciliation, through reminders of how what may seem abrasive may be a difference in cultural language around distress. Future work could leverage posts in newly designed spaces (building on our analytic approach) to analyze how reconciliatory language across partisans takes form and is shaped. This work could also analyze the role of design in these spaces by comparing reconcilatory language in designed spaces with reconciliatory language in threads from existing mental health subreddits that have both Republican and Democrat forum participants interacting with each other.

For example, in the mental health context, online platforms could remind users that culture often influences how people express themselves, and while polarizing language may be indicative of a troll~\cite{simchon2022troll}, it may also be indicative of a person who is hurting and needs support. Helping users to be more conscious of the diverse ways that distress is expressed, including alongside polarizing or stigmatizing language, may pave the way for users to more deeply listen to the person on the other side of the screen, and form bonds over shared distress. Platform designers and community moderators can actively promote conscientiousness and bond formation by implementing guidelines that encourage respectful dialogue, providing resources that support understanding between people from different political backgrounds, and facilitating structured discussions that foster empathy among users who express their distress differently. 

Bonds between Republican and Democrat users established on these platforms could help support bipartisan movements towards mental health policy reform, led by people with lived experience. Advocates could emphasize that, though the language that Republicans and Democrats utilize to express their mental health may be different, the distress they are experiencing is similar, and can be alleviated through greater policy initiatives. Past work has demonstrated that state mental health policies are often reflective of the dominant party's worldview on mental health~\cite{conley2023predictors}. Exposing partisan users to the notion of diverse expressions of distress may allow for shared bonds. 

\subsection{Limitations and Future Work}
We operationalized partisan culture as a dimension to understand mental health expressions, scoped to partisan culture within the U.S. Future work could study additional dimensions of social, political, partisan, and ideological identity outside of the U.S., and make connections across borders. For example, a question that could build on this work is understanding whether ideological alignment (such as more conservative or liberal political beliefs) has an influence on expressions of distress, and whether that changes based on national context across countries. 

Our dataset is comprised of people who chose to tie their identity in partisan subreddits to mental health subreddits, or people who vocally and actively associate with a given partisan culture and with their mental health experiences. This is a particularly unique set of users, given that some users will utilize anonymous \textit{throwaway} accounts when posting about their mental health on Reddit~\cite{de2014mental}, and often only post once when doing so~\cite{pavalanathan2015identity}. Our dataset is also comprised of people who are Reddit users, who are not representative of the national population of Republicans and Democrats with lived experience of mental illness at large. This is particularly important to acknowledge given that many Republicans live in rural areas~\cite{parker2018urban}, in which approximately 30\% of people do not have Internet connectivity or access~\cite{vogels2021some}. Our dataset consists of a matched sample of 8,916 Republican, Democrat, and Unaffiliated users, with a 1-1 match for every Republican user identified in our dataset, given that there are far fewer Republican users than Democrat and Unaffiliated users. We utilize this approach to limit the influence of demographic and usage-based factors in the partisan differences we observe, but further speaking to representation, a limitation of this approach is that a considerable number of Democrat users are excluded from the analysis. Additionally, many users we dub Unaffiliated are strictly unaffiliated in an online context---they may very well associate with a partisan culture, but simply not assert it alongside their mental health experiences when posting on Reddit. Future work could leverage social media data donation (as has been done in mental health~\cite{razi2022instagram} and political~\cite{couto2024examining} contexts) of mental health posts from different online environments from a representative sample of Democrats and Republicans in the US, and use methods from political polling~\cite{Mercer2018Weighting} to better understand differences among a broader sample of Republican and Democrat users. Additionally, we do not derive causation between partisan cultural identity and distress language. An experimental design that isolates partisan affiliation and background as the key variable affecting the way mental health is described could test for causal differences. Such a study would require explicit disclosures of mental health and partisan affiliation, likely alongside other demographic attributes, which may be sensitive information. Even so, our findings provide insights into potential connections between partisan culture and mental health expression. 

We found that Democrat users tend to employ higher levels of clinical language in comparison to Republicans users. However, the underlying factors that influence this pattern are unclear---the phenomenon we observe may be the impact of lower mental health literacy among Republicans, but it also may be the impact of stigma, in which Republican users are aware of mental health language but choose to not use it out of shame. It also may be the case that Republican users choose not to use mental health language due to some other non-clinical or medical idiom of distress speaking more accurately to their experience of distress, or being more effective at permitting them access to care given the cultural dynamics of their area. It is also possible that psychiatric language has become an idiom of distress that is more commonly used by those from Democrat partisan cultural backgrounds for all kinds of distress. Foulkes~\cite{foulkes2023mental} has argued that people have become more likely to use psychiatric language to describe common distress over the course of the last decade---this phenomenon may be especially pronounced for Democrat users. Analyses done to track changes in Democrat and Republican partisan idioms of distress over time could help to answer this question. Additionally, interviews conducted with Republicans and Democrats with lived experience of mental illness could help shed light on the lived daily implications of some of the differences we observed.  

\section{Conclusion}
Our study presents how a ``personal is political''~\cite{guralnik2023couples} dynamic may manifest in online mental health discussions among Republican and Democrat partisan users. We find that Democrat users lean towards clinical and political polarization language in their expressions of distress, whereas Republican users lean towards language indicative of social relationships and contexts. We perceive these differences in expression not as obstacles, but as guideposts for the development of mental health approaches that bridge partisan divides and ensure individuals' comfort when accessing mental healthcare, despite highly stigmatized contexts. We argue that a greater consideration of partisan language around distress can support mutual understanding and collaborative action for better mental healthcare. In line with Gobodo-Madikizela's writings~\cite{gobodo2004human}, we discuss expressions of personal struggle with mental illness as a potential catalyst for coalition-building and healing. Through investigating the role of partisan culture in online mental health expressions, we shed light on the complex interplay between politics, culture, and lived experience of distress, paving the way for more effective approaches to mental health support and policy reform.


% \section*{Appendices}
% \subsection*{Appendix A: Unaffiliated Comparison, Expressive Differences}
% \repsunaffiliated
% \demsunaffiliated

% \subsection*{Appendix B: Unaffiliated Comparison, Clinical Language}
% \repsunaffiliatedclinical
% \demsunaffiliatedclinical

% \subsection*{Appendix C: Unaffiliated Comparison, Polarization Language}
% \repsunaffiliatedpolarization
% \demsunaffiliatedpolarization

\includepdf[pages=-]{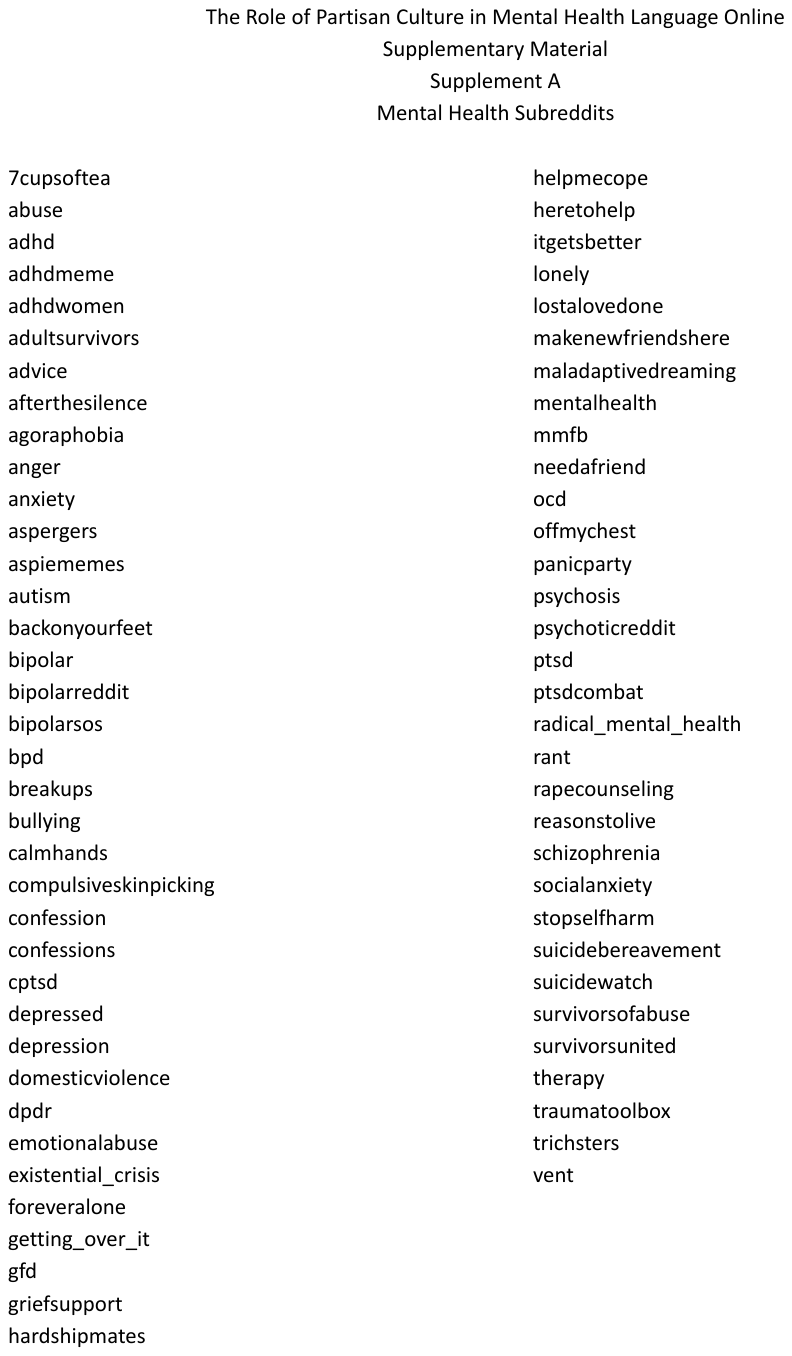}

\end{document}
\endinput
%%
%% End of file `sample-manuscript.tex'.